\documentclass[aps,pra,notitlepage,onecolumn,floatfix,superscriptaddress]{revtex4-1}
\usepackage{amsmath,amssymb,amsfonts,bbm,graphicx,color}
\usepackage[utf8]{inputenc}
\usepackage[T1]{fontenc}
\usepackage{physics}
\usepackage{times}
\usepackage{amsthm}
\usepackage[caption=false]{subfig}

\newcommand{\mar}[1]{{\color{black} #1}}

\usepackage{enumitem}   
\newcommand{\vecP}{\boldsymbol{\phi}}
\newcommand{\vecQ}{\mathbf{Q}}
\usepackage{booktabs}
\usepackage{float}
\usepackage{graphicx}
\usepackage{stmaryrd}
\usepackage{amsmath,amsfonts,amssymb}
\usepackage{multirow}
\usepackage{physics}
\usepackage{siunitx}
\usepackage{dsfont}
\usepackage{pgfplots}
\usepackage{wrapfig}
\usepackage{caption}	
\usepackage{bm}
\usepackage{soul}

\begin{document}

\title{Engineering dissipation with resistive elements in circuit quantum electrodynamics}

\author{Marco Cattaneo}
\affiliation{QTF Centre of Excellence, Department of Applied Physics, School of Science, Aalto
University, FI-00076 Aalto, Finland}
\affiliation{QTF Centre of Excellence, University of Helsinki, P.O. Box 43, FI-00014 Helsinki, Finland}
\affiliation{Instituto de Física Interdisciplinar y Sistemas Complejos IFISC (CSIC-UIB),
Campus Universitat Illes Balears, E-07122 Palma de Mallorca, Spain}
\author{Gheorghe Sorin Paraoanu}
\affiliation{QTF Centre of Excellence, Department of Applied Physics, School of 
Science, Aalto University, FI-00076 Aalto, Finland}

\begin{abstract}

The importance of dissipation engineering ranges from universal quantum computation to non-equilibrium quantum thermodynamics. In recent years, more and more theoretical and experimental studies have shown the relevance of this topic for circuit quantum electrodynamics, one of the major platforms in the race for a quantum computer. This article discusses how to simulate thermal baths by inserting resistive elements in networks of superconducting qubits. Apart from pedagogically reviewing the phenomenological and microscopic models of a resistor as thermal bath with Johnson-Nyquist noise, the paper introduces some new results in the weak coupling limit, showing that the most common examples of open quantum systems can be simulated through capacitively coupled superconducting qubits and resistors. The aim of the manuscript, written with a broad audience in mind, is to be both an instructive tutorial about how to derive and characterize the Hamiltonian of general dissipative superconducting circuits with capacitive coupling, and a review of the most relevant and topical theoretical and experimental works focused on resistive elements and dissipation engineering. 
\end{abstract}

\maketitle

\section{Introduction}
Circuit quantum electrodynamics (circuit QED) is nowadays at the forefront of experimental quantum physics, offering an experimental platform for topical research fields such as quantum information processing \cite{Wendin2016,Devoret2013,Arute2019}, quantum simulation \cite{Schmidt2013a,Paraoanu2014}, and quantum thermodynamics \cite{PekolaQT,Vinjanampathy2016,Silveri_2017}. By employing few circuit elements, namely capacitors, inductors, Josephson junctions and transmission line resonators, the experimentalist using this platform is able to initialize, measure and control the dynamics of one or more qubits, interacting or not with an external field \cite{Bouchiat1998,Blais2004}.  

The quantization of the above-mentioned circuit components is quite straightforward and well-understood, but including resistors in the formalism of circuit QED is not a trivial task. This is due to the fact that, while capacitors, inductors, junctions and finite transmission line resonators are non-dissipative elements modeleded by Hamiltonians, resistors induce energy loss in the circuits. This problem can be solved by adapting the phenomenological Caldeira-Leggett model of dissipation employing an infinite number of bosonic modes to describe resistive elements in circuit QED \cite{devoret1995}. When quantized through this method, a resistor behaves as a thermal bath \cite{BreuerPetruccione} with an Ohmic spectral density, reproducing the well-known Johnson-Nyquist (or thermal) noise \cite{Johnson1928,Nyquist1928a}. This idea is fundamental for the study of engineered dissipation in circuit QED. Our aim is to provide a broad introduction to this method, together with the necessary tools for its application in general scenarios, which may guide the reader through the study of networks of qubits and resistors in circuit QED. 

We start by presenting a pedagogical, step-by-step derivation of the Hamiltonian of the Caldeira-Leggett model and of the spectral density of the voltage fluctuations across a resistor. We do not assume for our presentation any previous knowledge about the topic, and close familiarity with quantum electromagnetic circuits is not strictly necessary throughout the paper, which we have written with a broad audience of theoretical and experimental quantum physicists in mind.

Having understood the principles of the Caldeira-Leggett model and how to apply them, it is important to clarify the physical intuition of how and why this phenomenological model works at all. Indeed, the phenomenological models for resistors do not depend on the specific materials used in their fabrication or on specific microscopic models; rather, they are constrained only by the general principles of thermodynamics, as first pointed out by Nyquist already in 1928 \cite{Nyquist1928a}. In order to understand the microscopic origin of fluctuations in resistors, we introduce the Landauer-Büttiker method for the study of scattering processes in a conductor as presented in the original papers \cite{Buttiker1990,Buttiker1992a,Buttiker1992,DeJong1994} and in a  well-known review on the topic \cite{Blanter2000}. We show that a description of the resistor based on this method provides a microscopic model that correctly reproduces the Johnson-Nyquist spectrum.

After introducing the phenomenological and microscopic models of resistors in circuit QED, we  focus on how to employ the latter as thermal baths to simulate open quantum systems \cite{BreuerPetruccione}. In order to do so, it is crucial to understand how the coupling between them and any circuit element (such as a transmon qubit \cite{Koch2007a}) works, and how to tune it. We  present the derivation of the general Hamiltonian of a circuit with capacitively coupled qubits and resistors, and discuss the master equations driving the open qubit evolution. Obtaining the final circuit Hamiltonian is not an easy task, and we  rely on a recent seminal paper that has paved the way for the treatment of any complex circuit network of coupled systems of this kind \cite{Parra-Rodriguez2018}. Furthermore, we  introduce some new results about the derivation in the weak coupling limit, leading to Markovian master equations. In particular, we  show how the reservoir engineering allows for the simulation of the open system dynamics of one or two qubits coupled to common or separate baths. The same procedure trivially applies for the study of harmonic oscillators (LC circuits in the present framework) immersed in thermal reservoirs. Moreover, we  briefly show how one can introduce a LC circuit acting as a filter, in order to shape the spectral density of the dissipation on the qubits.

To conclude the discussion, we review some of the major experimental methods to insert dissipative elements into superconducting circuits, with an eye to reservoir engineering. Fabricating resistive components in circuit QED is relatively easy, since they usually consist of simple normal-metal elements in a circuit made of superconducting materials \cite{Ronzani2018}. The dissipation induced by these elements is nowadays well-understood, as well as their thermodynamic properties, including the role played by the electron-phonon interaction \cite{Pothier1997b,Giazotto2006}. 
Resistors as independent elements of quantum circuits have been recently employed in disparate fields, with a particular focus on quantum thermodynamics \cite{PekolaQT}, and experiments relying on them have been presented in the past few years \cite{Gasparinetti2015,Partanen2016,Tan2017,Ronzani2018}. 
Besides resistors, transmission lines and lossy cavities have been used in circuit QED either as 3D or coplanar microwave components, and they can be modeled by similar techniques \cite{Murch2012,Shankar2013,Mlynek2015}.  In addition, we outline a few of the recent inroads into modern approaches on the active control of dissipation. While reducing decoherence is nowadays a major effort on the road to quantum computing, here we show, in contradistinction, that dissipation can be engineered to achieve stabilization of certain states in superconducting three-level systems and in resonators. A student in microwave engineering or rf circuits discovers quite early one little puzzle of this type: in a series RLC circuit, the quality factor is $(1/R)\sqrt{L/C}$, in agreement with our naive intuition that an increased resistance leads to  a worse quality factor for the oscillator. But for a parallel RLC circuit, the quality factor is $R\sqrt{C/L}$, increasing as $R$ gets larger -- the reason being, of course, that for large $R$ the current will flow less through the shunt resistor and more into the ideal $LC$ oscillator itself. The end of our paper highlights such situations where dissipation is beneficial, reviewing recent microwave engineering methods based on filtering, tunability, and microwave-assisted couplings to lossy transmission lines. Having at our disposal a controllable dissipative reservoir acting on superconducting qubits would be of great interest for both quantum engineers and quantum physicists \cite{Kapit2017}. The goal is to use dissipation as a resource for creating, in conjunction with coherent driving \cite{MULLER20121}, novel quantum states of matter and novel forms of quantum control. The benefit is stabilization and robustness, which comes for free.  For digital quantum computing, the usefulness of dissipation is mostly related to the realization of precise and fast initial states, and to some extend, if a modular-based approach is to be pursued, that of quantum memories \cite{Pastawski2011}.  However, alternative directions for quantum computing based on control of dissipation can be envisioned, where the steady-state negociated by the environment and coherent control encodes the result of a computation \cite{Verstraete2009}. These are forms of analog quantum processing of information \cite{RevModPhys.86.153,Paraoanu2014}. For simulating many-body systems, superconducting circuits provide time-resolved measurement data for each lattice site, allowing the extraction of higher-order correlations, which enable the investigation of thermalization processes and dynamics of phase transitions.

The paper is organized as follows: in Sec.~\ref{sec:model} we introduce the phenomenological and microscopic model of resistors in circuit QED. After presenting the concept of impedance and its extension for physical signals, we discuss the Caldeira-Leggett model for resistors and calculate the spectral density, showing its relation with the Johnson-Nyquist noise. Then, we introduce the Landauer-Büttiker method to sketch a microscopic model of the resistor reproducing the Johnson-Nyquist noise. After reviewing the capacitive coupling in Sec.~\ref{sec:capCoupling}, in Sec.~\ref{sec:coupling} we address how to employ it to connect resistors with other circuit elements, focusing in particular on the cases of one or two qubits, and discussing how to add a spectral filter. The weak coupling limit plays here a major role, while the general derivation for strong couplings is discussed in Appendix~\ref{sec:derivation}. Sec.~\ref{sec:control} addresses the main available control methods of dissipation, and reviews some experiments that make use of them. Finally, in Sec.~\ref{sec:conclusions} we briefly draw some concluding remarks and perspectives.
 
 In summary, our purpose is to establish a toolbox of theoretical models for dissipation engineering, with a focus on the comprehension of the Caldeira-Leggett model and its application in networks of coupled qubits. 

\section{A Hamiltonian formalism for resistors in circuit QED}
\label{sec:model}
In this section we  discuss how to describe a resistor in the time-independent Hamiltonian formalism for the quantization of quantum circuits, which the reader can find in Appendix~\ref{sec:quantization} with an application to the LC circuit. 
We introduce basic notions of circuit quantization and for further details we refer the reader to many excellent reviews about this topic \cite{Makhlin2001,Devoret2004,Wendin2007a,Clarke2008,Schoelkopf2008a,Devoret2013,Oliver2013,Xiang2013a,Schmidt2013a,Girvin2014,Paraoanu2014,Wendin2016,Gu2017,Vool2017,Kapit2017,Krantz2019,Nori2019,Huang2020,Kjaergaard2020,Blais2020,Rasmussen2021}.
After introducing the definition of impedance (or equivalently admittance) and its generalization for physical signals, we  focus on the Caldeira-Leggett model of a resistor presented in one of the very first lectures about quantized electromagnetic circuits \cite{devoret1995}, and we  follow the lines of this work to pedagogically review the model derivation. Next, we  calculate the spectral density of the voltage fluctuations across the resistor, and we show that it is equal to the one of a microscopic physical model based on the Landauer-Büttiker method.

\subsection{Impedance and admittance}
In the language of electrical engineering, the impedance of a circuit element is defined when dealing with AC signals.
Already at this initial step in the journey, the astute and diligent scholar will notice the discrepancy between the conventions used by physicists and engineers when it comes to defining Fourier transforms, phasors, and positive/negative frequencies. Throughout this paper, we adopt the definition employed in Ref. \cite{Clerk2010}, and we give below a number of considerations on how to relate this to the standard notations used in microwave engineering. 
Given a time-dependent operator or complex function $F (t)$, its Fourier transform is 
\begin{equation}
F (\omega ) = \int_{-\infty}^{\infty} dt  F(t) e^{i \omega t}    \label{eq:Fourier}
\end{equation}
with inverse
\begin{equation}
F(t) = \frac{1}{2\pi} \int_{-\infty}^{\infty} d\omega F(\omega)e^{-i\omega t}.
\end{equation}
If F is an operator, the direct and inverse Fourier transform of the adjoint are, by the definition above,
\begin{equation}
F^{\dag} (\omega ) = \int_{-\infty}^{\infty} dt  F^{\dag} (t) e^{i \omega t},    
\end{equation}
and similarly (with complex conjugation instead of the dagger symbol) for a complex function.
The following useful identities regarding a change of sign in frequency, hold true: $F(-\omega)= [F^{\dag}(\omega )]^{\dag}$ and
$F^{\dag}(-\omega)= [F(\omega )]^{\dag}$.
One notices that these definitions are generally at loggerheads with the ones used in elementary signal analysis. \mar{We provide additional clarifications about these conventions in Appendix~\ref{sec:sign}.}

We define the {\it complex impedance} $Z(\omega)$ of a branch of a circuit by
\begin{equation}
    Z(\omega ) = \frac{V(\omega )}{I(\omega )}.
    \label{eqn:impedDef}
\end{equation}
The same definition can be given in terms of the phasors $Ve^{-i \phi_{V}}$ and $Ie^{-i \phi_{I}}$ as $Z(\omega ) = (V/I) e^{-i (\phi_{V} - \phi_{I})}$. Since $V(t)$ and $I(t)$ are real functions, using the properties of the Fourier transform above we also have $V(-\omega )=[V(\omega )]^{*}$, $I(-\omega )=[I(\omega )]^{*}$,
therefore 
\begin{equation}
Z(-\omega )=[Z(\omega )]^{*}. 
\label{eq:zzstar}
\end{equation}
As mentioned already above, this relation allows us to translate our results with ease into the engineering convention if needed.


The impedance can be interpreted as the opposition to the passage of alternate current through the branch when a certain voltage is applied. 
Equivalently, we can use the \textit{admittance} $Y$ as a measure of how easily the current runs under the driving of an applied voltage:
\begin{equation}
\label{eqn:admittanceDef}
Y(\omega)=\frac{1}{Z(\omega)}.
\end{equation}

We can easily compute the impedance of all the passive circuit elements. Indeed, since Ohm's law for a resistor with resistance $R$ reads $V=RI$, we have
\begin{equation}
\label{eqn:impedanceRes}
Z_R(\omega)=R,
\end{equation}
being characterized by a constant value all over the range of frequencies. When dealing with a capacitor with capacitance $C$, we employ the relation $I=CdV/dt$ and obtain:
\begin{equation}
\label{eqn:impedanceCap}
Z_C(\omega)=\frac{1}{-i\omega C}.
\end{equation}
Finally, the constitutive equation for an inductor with inductance $L$ reads $V=LdI/dt$, therefore
\begin{equation}
\label{eqn:impedanceInd}
Z_L(\omega)=-i\omega L.
\end{equation}

From Eq.~\eqref{eqn:impedDef} we observe that the impedance of a series of circuit elements is given by the sum of the impedances of the elements. On the contrary, the inverse of the impedance of some circuit elements in parallel is given by the sum of the inverses of the impedances of the elements. Hence, the impedance of a parallel LC circuit reads
\begin{equation}
Z_{LC\parallel}(\omega)=\left(-i\omega C+\frac{i}{\omega L}\right)^{-1},
\end{equation}
which, after few steps, can be rewritten as
\begin{equation}
\label{eqn:impedanceLC}
Z_{LC\parallel}(\omega)=\frac{i}{2C}\left(\frac{1}{\omega+\omega_{LC}}+\frac{1}{\omega-\omega_{LC}}\right),
\end{equation}
where $\omega_{LC}=1/\sqrt{LC}$ is the resonant frequency of the circuit.

When discussing quantum electromagnetic circuits, we do not want to restrict ourselves to the AC circuit regime, but, on the contrary, we are interested in dealing with generic signals. Due to the linearity of the newtork (at least in the dissipative part), such signals, namely $V(t)$ and $I(t)$, can be written as a sum of the AC signals 
over all the frequencies, i.e. as a Fourier transform:
\begin{equation}
\label{eqn:realSignals}
V(t)=\frac{1}{2\pi}\int_{-\infty}^\infty d\omega\,V(\omega)e^{-i\omega t},
\end{equation} 
and equivalently for $I(t)$. In Eq.~\eqref{eqn:realSignals} we consider negative frequencies also, since $V(t)$ and $I(t)$ are real, physical signals, and the negative-frequency contribution deletes the imaginary part. Since $V(t)$ and $I(t)$ are indeed physical signals, they properly vanish at $t=\pm\infty$, therefore their Fourier transforms are well-defined.
Starting from Eq.~\eqref{eqn:impedDef}, we write the relation between voltage and current by making use of a linear response function: 
\begin{equation}
\label{eqn:linearResponseImpedance}
V(t)=\int_{-\infty}^\infty dt'\,Z(t-t')I(t').
\end{equation}
where $Z(t)$ is the Fourier transform of the complex impedance $Z(\omega)$. Some divergence issue may arise in the last step; it turns out that, to avoid any problem, we need to employ a \textit{generalized impedance} function, defined as:
\begin{equation}
\label{eqn:generalizedImpedance}
Z^g(\omega)=\lim_{\eta\rightarrow 0^+} Z(\omega+i\eta)=\lim_{\eta\rightarrow 0^+}\int_{-\infty}^\infty dt\,Z(t)e^{-\eta t}e^{i\omega t}.
\end{equation}
Using Eq.~\eqref{eqn:generalizedImpedance}, the relation between current and voltage reads 
\begin{equation}
\label{eqn:linearResponseImpedanceGen}
V(t)=\lim_{\eta\rightarrow 0^+}\int_{-\infty}^\infty dt'\,Z(t-t')e^{-\eta (t-t')}I(t').
\end{equation}
In Appendix~\ref{sec:linearResp} we carefully discuss the derivation of Eq.~\eqref{eqn:linearResponseImpedance} and the meaning of Eq.~\eqref{eqn:generalizedImpedance}. Here, we just point out that such definition is necessary to introduce the linear response function of the LC circuit, given that Eq.~\eqref{eqn:impedanceLC} does not admit Fourier transform because of the poles at $\omega=\pm \omega_{LC}$. In this case, employing Eqs.~\eqref{eqn:generalizedImpedance} and~\eqref{eqn:linearResponseImpedanceGen} corresponds to shifting the poles in the lower half plane. Moreover, notice that the negative exponential in Eq.~\eqref{eqn:generalizedImpedance} does not cause any divergence issue, since for $t<0$ causality requires $Z(t)=0$ \cite{Jackson}. Finally, the definition of a linear response function as in Eq.~\eqref{eqn:linearResponseImpedance} implies a couple of relations between the real and imaginary part of $Z(\omega)$, known as Kramers-Kronig relations \cite{Jackson}:
\begin{equation}
\label{eqn:KramersKronig}
\begin{split}
&\Re[Z(\omega)]=\frac{1}{\pi}\mathcal{P}\int_{-\infty}^\infty d\omega'\,\frac{\Im[Z(\omega)]}{\omega'-\omega},\\
&\Im[Z(\omega)]=-\frac{1}{\pi}\mathcal{P}\int_{-\infty}^\infty d\omega'\,\frac{\Re[Z(\omega)]}{\omega'-\omega},
\end{split}
\end{equation}
where $\mathcal{P}$ denotes the principal value integral.

\subsection{Caldeira-Leggett phenomenological model of a resistor}
\label{sec:Devoret}

A phenomenological model for a resistor can be obtained by adapting the seminal concepts introduced by Caldeira and Leggett in 1983 to describe dissipation in quantum mechanics \cite{Caldeira1983}. The key idea is to describe the environment by a collection of harmonic oscillators - which, in the case of electrical circuits, can be taken as a collection of LC oscillators. Such a model was formulated with clarity and in modern notations and electrical-engineering terminology by Michel Devoret in the lecture notes of \textit{Les Houches} summer school in 1995 \cite{devoret1995}, and further refined in  Ref.~\cite{Vool2017}.  The model is based on a method of network synthesis used to represent, under certain assumptions, a generic purely imaginary impedance as a collection of LC circuits. Different ways of realizing such synthesis are available, and here we focus on the ones called first and second Foster's form \cite{Foster1924,Smith1992}, extending the method to dissipative impedances, characterized by a real part also. We mention here that a renewed interest in the description of a generic impedance (or admittance) in circuit QED has emerged in the last twenty years, and has led to the development of novel theoretical methods. For instance, a couple of seminal papers by Burkard et al. addressed the relaxation and decoherence times of superconducting flux and charge qubits in the presence of an environment, where the latter is described as a collection of harmonic oscillators in a Caldeira-Leggett fashion, considering weak coupling and possible leakages to higher qubit levels \cite{Burkard2004,Burkard2005}. \textit{Black box quantization}, presented in Ref.~\cite{Nigg2012}, is a method of obtaining the effective low energy Hamiltonian of one (or more) Josephson junctions in parallel with a generic environment, by describing the latter through the Foster's first form. A refined version of this method, relying on Brune's synthesis instead of Foster's one, has been presented in Ref.~\cite{Solgun2014} and later extended to multiport impedances \cite{Solgun2015}. The well-established input-output theory to compute dissipation \cite{Yurke1984} has been broadly employed since the very early stages of circuit QED, with a special attention on the analysis of open transmission line resonators with an infinite number of modes \cite{Leppakangas2018}. \mar{Quantization of generic networks in circuit quantum electrodynamics has been extensively analysed by Parra-Rodriguez et al. \cite{Parra-Rodriguez2021}, considering both reciprocal \cite{Parra-Rodriguez2018} and non-reciprocal \cite{Parra-Rodriguez2019} devices.} Finally, a recent paper proposes a new method to derive the master equation of a system of superconducting qubits coupled to a low admittance, corresponding to weak coupling to the environment, by extracting only a few relevant degrees of freedom of the circuit and treating the rest in a Lindbladian way \cite{Hassler2018}.

\subsubsection{Building the model}
\begin{figure}
\center
\includegraphics[scale=0.5]{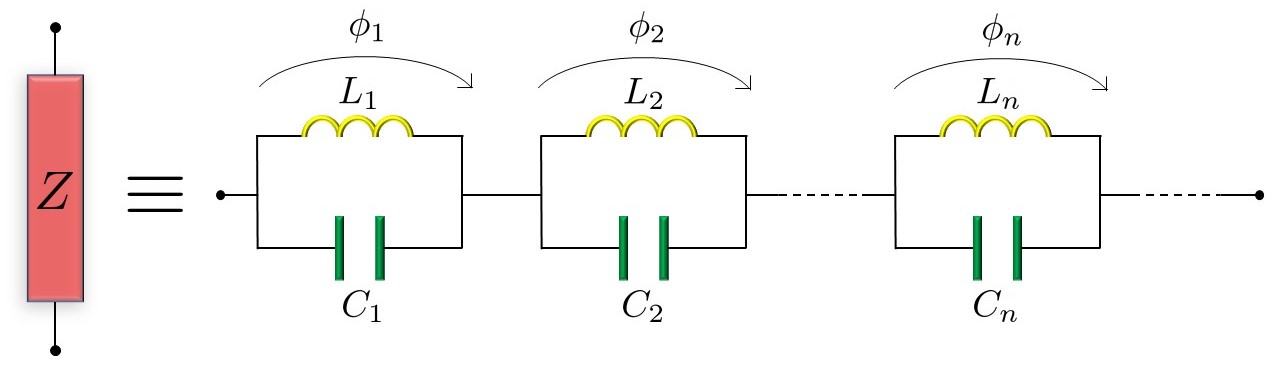}
\caption{Foster's first form to describe an impedance as an infinite series of parallel LC circuits. For convenience, we define the node fluxes of the system as in the figure above, by considering the voltage across each LC circuit.}
\label{fig:firstFoster}
\end{figure}
Let us start with the Foster's first form to describe an impedance, say $Z(\omega)$, by employing capacitors and inductors only. We consider a network composed of a collection of parallel LC circuits in series, as in Fig.~\ref{fig:firstFoster}. If the number of parallel LC circuits is finite, the overall impedance of the network must be purely imaginary. On the contrary, if we consider an infinite number of them, the real part of the impedance appears. Indeed, by making use of Eq.~\eqref{eqn:impedanceLC} and of the law on the impedance of a series of elements, the generalized impedance of the network in Fig.~\ref{fig:firstFoster}, which we name as $Z_\infty(\omega)$, reads
\begin{equation}
\label{eqn:sumDevoret}
Z_\infty(\omega)=\lim_{\eta\rightarrow 0^+}\sum_{j=1}^\infty \frac{i}{2C_j}\left(\frac{1}{\omega-\omega_j+i\eta}+\frac{1}{\omega+\omega_j+i\eta}\right),
\end{equation}
where $C_j$ and $\omega_j$ are respectively the capacitance and the resonant frequency of the $j$th LC circuit.
Now, we carefully choose the values of each $C_j$ and $\omega_j$, such that:
\begin{equation}
\label{eqn:valueDevoretComega}
C_j=\frac{\pi}{2\Delta\omega\Re[Z(j\Delta\omega)]},\qquad \omega_j=j\Delta\omega,
\end{equation}
where $Z(\omega)$ is the impedance whose action we want to mimic and $\Delta\omega$ is very small. The corresponding inductance $L_j$ of each LC circuit is readily obtained as:
\begin{equation}
\label{eqn:valueDevoretL}
L_j=\frac{2\Delta\omega\Re[Z(j\Delta\omega)]}{\pi\omega_j^2}.
\end{equation}

If $\Delta\omega$ is small enough to be considered as infinitesimal, the infinite summation in Eq.~\eqref{eqn:valueDevoretComega} can be seen as an integral over the (now continuous) variable $\omega_j$, and after inserting the chosen values we obtain:
\begin{equation}
Z_\infty(\omega)=\lim_{\eta\rightarrow 0^+}\int_0^\infty d\omega_j\,\frac{i}{\pi}\Re[Z(\omega_j)]\left(\frac{1}{\omega-\omega_j+i\eta}+\frac{1}{\omega+\omega_j+i\eta}\right).
\end{equation}
We can now apply the Plemelj-Sokhotski theorem for the real line\footnote{Stating that $\lim_{\eta\rightarrow 0^+}\int_a^b dx\,\frac{f(x)}{x\pm i\eta}=	\mp i\pi f(0)+\mathcal{P}\int_a^b dx\,\frac{f(x)}{x}$, if $a<0<b$ and $f(x)$ is defined and continuous on the integration interval.} to the above integral, splitting the latter into a real and an imaginary part. Notice that the presence of a real part is due to the infinite number of LC circuits characterized by dense values of the resonant frequency. \mar{Indeed, a non-zero real part, which will be proportional to the resistance $R$, indicates the presence of dissipation in the circuit. From a physical perspective, its origin can be traced back to the \textit{infinite} number of LC circuits: intuitively, an input signal can be sent through the chain of LC circuits and, since it will never reach the ``end'' of the chain, it will never ``bounce back'' and return to the start. This means that input excitations can travel ``forever'' through the LC chain without any recurrence time, that is, energy at the input is dissipated. This heuristic picture shows how the infinite-dimensional Foster's decomposition is able to simulate physical dissipation by employing non-dissipative elements only, provided we use an infinite number of them.} Physically, the poles of the impedance at the resonance frequencies $\omega_{j}$ correspond to the resonant excitation of the $L_{j}C_{j}$ oscillator when the circuit is open-ended. 
 By studying separately the case $\omega>0$ and $\omega<0$, we obtain that the real and imaginary parts of $Z_\infty(\omega)$ are given by:
\begin{subequations}
\label{eqn:Zinf}
\begin{align}
\begin{split}
\Re[Z_\infty(\omega)]=\Re[Z(\omega)],
\end{split}\\
\begin{split}
\label{eqn:22b}
\Im[Z_\infty(\omega)]=\frac{1}{\pi}\left(\mathcal{P}\int_0^\infty d\omega_j \frac{\Re[Z(\omega_j)]}{\omega-\omega_j}+\mathcal{P}\int_0^\infty d\omega_j \frac{\Re[Z(\omega_j)]}{\omega+\omega_j}\right).
\end{split}
\end{align}
\end{subequations}
We have been successful in reproducing the real part of the impedance $Z(\omega)$ by employing an infinite network of passive capacitors and inductors only. The relation between real and imaginary part given by Eq.~\eqref{eqn:Zinf} expresses the Kramers-Kronig relations in Eq.~\eqref{eqn:KramersKronig}, stressing that voltage and current depend on each other through a linear response function. 

A suitable expression for the impedance of a resistor fulfilling Eq.~\eqref{eqn:Zinf} is given by:
\begin{equation}
\label{eqn:impedanceOhmic}
Z_R^{\textnormal{eff}}(\omega)=R\frac{\omega_C^2}{\omega_C^2+\omega^2}+i R\frac{\omega\omega_C}{\omega_C^2+\omega^2},
\end{equation}
where $\omega_C$ is a suitable cutoff frequency, and the above equation is valid in the range $\omega\ll \omega_C$. \mar{For a discussion about the cutoff frequency in circuit quantum electrodynamics, we refer the reader to Refs.~\cite{Malekakhlagh2017a,Gely2017a}.} Eq.~\eqref{eqn:impedanceOhmic} can be derived through the classical Lorentz-Drude dispersion model \cite{Jackson}, and is usually referred to as the \textit{Ohmic spectrum of a resistor}. The reader can verify that it fulfils the condition of Eq.~\eqref{eqn:Zinf} by substituting the value of $Z(\omega)$ in Eq.~\eqref{eqn:Zinf} with $Z_R^{\textnormal{eff}}(\omega)$ in Eq.~\eqref{eqn:impedanceOhmic}, and solving the integrals in Eq.~\eqref{eqn:22b}.

In the derivation of Eq.~\eqref{eqn:Zinf} we have discussed the cases $\omega>0$ and $\omega<0$, but not the value $\omega=0$. Indeed, from Eq.~\eqref{eqn:sumDevoret} we observe that, if $\omega=0$, then the impedance vanishes. This is a requirement of the Foster's first form, and it can be interpreted as the fact that, for a DC signal, the capacitors in Fig.~\ref{fig:firstFoster} do not allow for the passage of current. Therefore, the latter flows through each inductor, which does not oppose any resistance, i.e. the impedance is zero. This is a pathological case appearing only for the singularity $\omega=0$, therefore, if we deal with a physical signal whose contribution given by frequency $\omega=0$ (i.e. DC current) is negligible, we should not bother about it. However, if we want to treat the response of the system to DC current, we need to add an initial capacitor or inductor, depending on the behavior we want to mimic \cite{devoret1995,Vool2017}.

\subsubsection{Quantization of the resistor}
Let us now address the quantization of the resistor. We consider here the open circuit of Fig.~\ref{fig:firstFoster}, i.e. the two terminals of the resistor are grounded. In this case, we can parametrize the system as shown in Fig.~\ref{fig:firstFoster}, where we term the flux associated to the voltage difference between the $j$th LC circuit as $\phi_j$. Using this convention, we can see the LC circuits as a collection of decoupled harmonic oscillators, and we quantize the system through the standard quantization rules of circuit QED discussed in Appendix~\ref{sec:quantization}. The resistor Hamiltonian is readily written:
\begin{equation}
\label{eqn:HamiltonianResistor}
\mathcal{H}_R=\sum_{j=1}^\infty \left[\frac{Q_j^2}{2C_j}+\frac{\phi_j^2}{2L_j}\right]=\sum_{j=1}^\infty \hbar\omega_j a_j^\dagger a_j,
\end{equation}
where $Q_J=C_j\dot{\phi}_j$, $a_j^\dagger$ and $a_j$ are the standard creation and annihilation operators \cite{MandelWolf} and the values of $C_j,L_j,\omega_j$ are given by Eq.~\eqref{eqn:valueDevoretComega} and Eq.~\eqref{eqn:valueDevoretL}, while the resistor impedance thereof can be found in Eq.~\eqref{eqn:impedanceOhmic}. Note that, using the standard convention of quantum optics, we have neglected an infinite constant term in the right-hand side of Eq.~\eqref{eqn:HamiltonianResistor}.

To conclude the exposition of the Caldeira-Leggett model for a resistor, note that all the previous results can be obtained for an admittance instead of an impedance, by describing the equivalent network using the Foster's second form \cite{Vool2017}. All the equations of the present section still hold, switching voltage with current and capacitance with inductance. In this case, the poles of the admittance at $\omega_{j}$ correspond to the resonant excitation of the corresponding $L_{j}C_{j}$ oscillator, as the input of the line is shortcut.
The only difference consists in the fact that Foster's second form requires a vanishing admittance at $\omega=0$, that is to say, infinite impedance.

\subsubsection{Thermal noise}
We are now able to calculate the spectrum of the voltage fluctuations over the resistor, if the latter is in a thermal state at temperature $T=1/\beta k_B$. The thermal state is defined as:
\begin{equation}
\label{eqn:thermalStateMultimode}
\rho_B(\beta)=\frac{1}{\mathcal{Z}}e^{-\beta\sum_j\hbar\omega_j a_j^\dagger a_j}=\bigotimes_j \rho_B(\beta,\omega_j), 
\end{equation}
where $\rho_B(\beta,\omega_j)$ describes the thermal state of each mode:
\begin{equation}
\label{eqn:thermalStateSinglemode}
\rho_B(\beta,\omega_j)=\frac{1}{1+n(\omega_j)}\sum_{n=0}^\infty \left(\frac{n(\omega_j)}{1+n(\omega_j)}\right)^n\ket{n}_j\!\bra{n},
\end{equation}
where $\ket{n}_j$ are the eigenstates of the number operator of the mode $j$, $n_j=a_j^\dagger a_j$, and $n(\omega_j)=\langle n_j\rangle_{\rho_B(\beta,\omega_j)}=1/(e^{\hbar\omega_j\beta}-1)$ is the Bose-Einstein distribution. Fore more details about the thermal state we refer the reader to Ref.~\cite{MandelWolf}.

Let us now calculate the \textit{correlation function} of the voltage across the $j$th LC circuit. Such function is defined as
\begin{equation}
\label{eqn:correlationFunction}
C_{VV}^{(j)}(t)=\langle \dot{\phi}_j(t)\dot{\phi}_j(0)\rangle_{\rho_B(\beta)},
\end{equation}
where the evolution of any operator is computed according to the Hamiltonian Eq.~\eqref{eqn:HamiltonianResistor}, and we are exploiting its time-translation invariance to define the correlation function as depending only on $t$. From standard Hamilton's equations, we observe that $\dot{\phi}_j(t)=\dot{\phi}_j(0)\cos\omega_j t-\omega_j\phi_j(0)\sin\omega_j t $. We recall that $\phi_j(0)$ and $\dot{\phi}_j(0)$ are related to the creation and annihilation operators through:
\begin{equation}
\begin{split}
\phi_j(0)=\sqrt{\frac{\hbar}{2}\frac{1}{C_j\omega_j}}\left(a_j^\dagger+a_j\right),	\qquad
\dot{\phi}_j(0)=i\sqrt{\frac{\hbar}{2}\frac{\omega_j}{C_j}}\left(a_j^\dagger-a_j\right).
\end{split}
\end{equation}
Therefore,
\begin{equation}
\begin{split}
\langle \dot{\phi}_j(0)^2\rangle&=-\frac{\hbar\omega_j}{2C_j}\Tr[\rho_B(\beta)\left(a_j^\dagger-a_j\right)\left(a_j^\dagger-a_j\right)]=(2n(\omega_j)+1)\frac{\hbar\omega_j}{2C_j},
\end{split}
\end{equation}
where we have used Eq.~\eqref{eqn:thermalStateSinglemode} and the commutator $[a_j,a_j^\dagger]=1$. 
Analogously, we have
\begin{equation}
\begin{split}
\langle\phi_j(0)\dot{\phi}_j(0)\rangle&=i\frac{\hbar}{2C_j}\Tr[\rho_B(\beta)\left(a_j^\dagger+a_j\right)\left(a_j^\dagger-a_j\right)]=i\frac{\hbar}{2C_j}.
\end{split}
\end{equation}
Finally, we calculate the correlation function:
\begin{equation}
\label{eqn:correlationFunctionResistor}
\begin{split}
C_{VV}^{(j)}(t)&=(2n(\omega_j)+1)\frac{\hbar\omega_j}{2C_j}\cos\omega_j t-i\frac{\hbar\omega_j}{2C_j}\sin\omega_j t=\frac{\hbar\omega_j}{2C_j}\left[(n(\omega_j)+1)e^{-i\omega_j t}+n(\omega_j) e^{i\omega_j t}\right]. 
\end{split}
\end{equation}

The operator describing the voltage difference across the resistor is nothing but the sum of the voltages over each LC circuit, $V=\sum_j\dot{\phi}_j$. Since all the harmonic oscillators in Eq.~\eqref{eqn:HamiltonianResistor} are uncorrelated, after substituting the value of $C_j$ given by Eq.~\eqref{eqn:valueDevoretComega}, the correlation function of the voltage operator reads
\begin{equation}
C_{VV}(t)=\int_{-\infty}^\infty d\omega_j\,\frac{\hbar\omega_j}{\pi}\Re[Z(\omega_j)]\mathcal{N}(\omega_j)e^{-i\omega_j t},
\end{equation} 
where we have replaced the summation over all the modes with an integral, and we have introduced the notation $\mathcal{N}(\omega_j)= (n(\omega_j )+1) \Theta (\omega_j ) - n(-\omega_j )\Theta (-\omega_j )$. Here $n(\omega )=(e^{\beta \hbar \omega}-1)^{-1}$ and $\Theta(\omega)$ is the Heaviside theta function. From the result Eq. (\ref{eqn:correlationFunctionResistor}) above it is clear that positive frequencies correspond to absorption of energy and negative frequencies to emission. This can be seen by considering $T=0$ (vacuum state). In  this case only the positive-frequency part remains (indicating that the system is still able to absorb energy), while the negative-energy part is zero (indicating that it is not possible to emit the energy of vacuum). This is reflected in the definition of the quantity $\mathcal{N}(\omega )$, which is $\mathcal{N}(\omega ) = n(\omega ) + 1$ for absorptive processes ($\omega > 0$) and $\mathcal{N}(\omega ) = - n (-\omega )$ for emission ($\omega < 0$).


The \textit{spectral density} of an operator $O$ is defined 
as the Fourier transform of its correlation function \cite{Clerk2010}:
\begin{equation}
\label{eqn:spectrumDef}
S_{OO}(\omega)=\int_{-\infty}^\infty dt\,C_{OO}(t)e^{i\omega t}.
\end{equation}  

At this point it is useful to mention the connection with the Wiener-Khinchin theorem in signal analysis. Given a random process $X(t)$, the power spectral density (PSD) is defined as 
\begin{equation}
S_{XX}(\omega ) = \lim_{\mar{t_W}\rightarrow \infty} \frac{\langle |X_{\mar{t_W}}(\omega )|^2 \rangle}{\mar{t_W}} \label{eq:PSD}
\end{equation}
with 
\begin{equation}
    X_{\mar{t_W}}(\omega ) = \int_{-\mar{t_W}/2}^{\mar{t_W}/2} dt X(t)e^{i\omega t}
\end{equation}
being the windowed Fourier transform and $\langle ... ,... \rangle$ the ensemble average -- i.e. expectation value with respect to the probability distribution function (PDF) of the process. The quantity $|X_{\mar{t_W}}(\omega )|^2/\mar{t_W}$ is referred to as periodogram or sample spectrum. Suppose we record the signal with an oscilloscope in a time-window $\mar{t_W}$. We can take the Fourier tranform of the data and calculate $|X_{\mar{t_W}}(\omega )|^2/\mar{t_W}$, which is an unbiased estimator of the true power spectral density Eq. (\ref{eq:PSD}) obtained by averaging over an ensemble of such measurements.
For wide-sense stationary processes, the autocorrelation function depends only on the lag between the two times:
\begin{equation}
    C_{XX} (\tau ) = \langle X(t + \tau) X(t)\rangle.
\end{equation}
Then, the Wiener-Khinchin theorem says that
\begin{equation}
S_{XX}(\omega ) = \int_{-\infty}^{\infty}d\tau C_{XX}(\tau )e^{i\omega \tau}.\label{eq:WK}
\end{equation}
Hence, the power spectral density is the Fourier transform of the auto-correlation function, and following the reasoning above Eq.~\eqref{eqn:spectrumDef} can be interpreted as a mathematical result instead of a definition. For a detailed proof of this result we refer to Ref. \cite{Lu2009}

Finally, the spectral density of the fluctuations of the voltage operator across a resistor, reads
\begin{equation}
\label{eqn:spectrumResistor}
\begin{split}
S_{VV}(\omega)&=\int_{-\infty}^\infty d\omega_j \int_{-\infty}^\infty dt\,\frac{\hbar\omega_j}{\pi}\Re[Z(\omega_j)]\mathcal{N}(\omega_j)e^{-i(\omega_j-\omega) t}\\
&=2\hbar\omega\Re[Z(\omega)]\mathcal{N}(\omega)=\hbar\omega\left(\coth\frac{\beta\hbar\omega}{2}+1\right)\Re[Z(\omega)] = \frac{2 \Re[Z(\omega)]\hbar\omega}{1-e^{-\beta \hbar\omega}}.
\end{split}
\end{equation}

Eq.~\eqref{eqn:spectrumResistor} represents the quantum limit of the well-known Johnson-Nyquist noise induced by thermal agitation \cite{Johnson1928,Nyquist1928a}; we will see throughout the paper how this expression can be interpreted as the standard fluctuation-dissipation theorem for electrical circuits. For low frequencies or high temperatures we find the classical result $S_{VV}(\omega)=2k_B T\Re[Z(\omega)]$, with a factor $2$ instead of a factor $4$ because of the employed definition of spectral density including negative frequencies also (see the discussion below). This expression is referred to as the double-sided spectral density, since $\omega$ can take both positive and negative values.

The zero-voltage noise due to the thermal agitation of electric charges in a conductor was first observed experimentally by John B. Johnson \cite{Johnson1928}, and then described theoretically by Harry Nyquist \cite{Nyquist1928a} during 1926-1928, while both scientists were working at Bell Labs. Nyquist's derivation is a beautiful demonstration of the heuristic power of fundamental concepts from  thermodynamics and statistical mechanics. In his seminal paper from 1928, he considered two resistors with resistance $R$ at the same temperature $T$, connected by conductive wires. Since the resistors are in thermal equilibrium and have the same resistance, the electrical power due to thermal agitation flowing towards a resistor or the other must be same. This must hold true not only for the total power, but also the power exchanged within any frequency bandwidth. If this were not the case, by inserting a filter or a resonant circuit in-between the resistors one obtains immediately a contradiction with the second law of thermodynamics, as such a circuit would allow extracting work from a single reservoir at temperature $T$. It then follows that the voltage across the resistor 
must be a function of frequency, resistance and temperature only. 
Next, to derive the final formula, Nyquist analysed a circuit where two resistors are connected by a transmission line of length $l$.
Let us first calculate the power transmitted from one resistor to the other. If $V$ is the voltage generated by thermal noise corresponding say to the left resistor, then this creates a voltage $V R/(R+R) = V/2$ on the right resistor, and the corresponding power transmitted through the line to be dissipated in that resistor is $P=V^{2}/(4R)$. 
Next, we recall Parseval's theorem,
\begin{equation}
    \int_{-\infty}^{\infty}dt V^{2}(t) = \frac{1}{2\pi}\int_{-\infty}^{\infty}d\omega |V(\omega)|^{2}
\end{equation}
showing that the energy content of the signal is the same irrespective to whether it is expressed in time-domain or in frequency domain.  


Further, let us define the 
time-average power of the signal $V(t)$ by
\begin{equation}
\overline{V^2} = \lim_{\mar{t_W}\rightarrow \infty}\frac{1}{\mar{t_W}}\int_{-\mar{t_W}/2}^{\mar{t_W}/2}dt V^{2}(t). 
\end{equation}
We can now use Parseval's theorem in the relation above and, further, we can move the limit inside the integral by taking the ensemble average. But since we assume that the process is ergodic, the ensemble average should give the same result as the temporal average. Thus we obtain the mean squared voltage
\begin{equation}
\langle V^2 \rangle = \frac{1}{2\pi}\int_{-\infty}^{\infty} d\omega S_{VV}(\omega ), 
\end{equation}
where $S_{VV} (\omega ) = \lim_{\mar{t_W}\rightarrow \infty} \langle |V(\omega)|^{2}\rangle /\mar{t_W} $ is the voltage power spectral density, with $\langle ... , ... \rangle$ denoting ensemble average.

This means that, in the frequency space, the power is distributed over frequencies and therefore the quantity
\begin{equation}
\frac{1}{2\pi} S_{VV} (\omega ) d\omega
\end{equation}
represents the average voltage power in an interval $d \omega$ at a frequency $\omega$. This can be understood also as the result of filtering the voltage $V(t)$ with a bandpass filter centered at $\omega$ with bandwidth $d\omega$.

The same result can be obtained by applying the Wiener-Khinchin theorem. From Eq. (\ref{eq:WK}) we have 
$R_{XX}(\tau ) = \int_{-\infty}^{\infty} d\omega/(2 \pi) S_{XX}(\omega ) e^{-i\omega \tau}$ and for $\tau =0$ the autocorrelation function becomes simply the power, $R_{XX}(t) = \langle X^{2}(t)\rangle$. Identifying $X=V$ we obtain
\begin{equation}
\langle V^2 \rangle = \frac{1}{2\pi} \int_{-\infty}^{\infty} d\omega S_{VV}(\omega ).
\end{equation}

We can also see from the above expression that, when restricting the integrals to only positive frequencies, 
\begin{equation}
\langle V^2 \rangle = \frac{1}{2\pi} \int_{0}^{\infty} d\omega [S_{VV}(\omega ) + S_{VV}(-\omega )] =
\frac{1}{2\pi} \int_{0}^{\infty} d\omega \mathbb{S}_{VV}(\omega ). \label{eq:positiveonly}
\end{equation}
where the quantity  $\mathbb{S}_{VV}(\omega )= S_{VV}(\omega ) + S_{VV}(-\omega )$ is called the symmetrized single-sided spectral density, and has the meaning of voltage power per bandwidth at positive frequencies. 
Typical laboratory instruments such as spectrum analysers measure directly $\mathbb{S}_{VV}(\omega )$,  since they use
a local oscillator with positive frequencies to mix, downconvert, and power-detect the signal.

Returning now to our circuit, let us consider the positive-frequency case $\omega > 0$ as well. Since $V(t)$ is real, as we have seen in Section 2.1, we have $V(-\omega ) = [V(\omega )]^{*}$ and therefore classically $S_{VV}(\omega ) = S_{VV }(-\omega )$ and $\mathbb{S}_{VV}(\omega ) = 2 S_{VV} (\omega )$.

Next, we can  write the average power dissipated in the resistor in a frequency interval $d\omega$ as 
\begin{equation}
\frac{1}{2\pi} \frac{ \mathbb{S}_{VV}(\omega)}{4R} d\omega.
\end{equation}
 Therefore, the  energy $dE$ in this bandwidth can be obtained from the power above, transmitted through the transmission line during the time $l/v$ (the time needed for photons to traverse the line) by the two sources (the left and right resistors)

\begin{equation}
dE = \frac{l}{\pi v}\frac{\mathbb{S}_{VV}(\omega)}{4R} d\omega \label{eq:denergy1}.
\end{equation}
On the other hand, the energy $dE$ can be obtained by the following reasoning. After the equilibrium has been established, the transmission line is suddenly disconnected and shorted at both ends. This ensures that the energy is stored in modes satisfying $l = n \lambda /2$, where $\lambda = 2\pi v/\omega$ is the wavelength, $v$ is the speed of light in the line, and $\omega /(2\pi)$ is the frequency. Therefore the number of  modes in the frequency interval $d \omega/(2\pi)$ is $dn = ld \omega /( \pi v)$. From the equipartition theorem, we have to ascribe an energy $k_{\rm B}T$ per mode (each mode has two degrees of freedom, magnetic and electric). As a result,
\begin{equation}
    dE = k_{\rm B}T dn = \frac{l}{\pi v}k_{\rm B}T d\omega .\label{eq:equip}
\end{equation}
By comparing Eq. (\ref{eq:denergy1}) with Eq. (\ref{eq:equip}) we find the famous formula \cite{Nyquist1928a}
\begin{equation}
\mathbb{S}_{VV} = 4 k_B T R, \label{eq:famous}
\end{equation}
which is valid in the large-temperature limit $k_{\rm B} T \gg \hbar \omega$. Nyquist also showed how to generalize this formula for any complex impedance, which amounts in our model to the replacement of R by $\mathrm{Re}[Z(\omega )]$. To cover also the large-frequency limit, he proposed to use Planck's distribution instead of the equipartition theorem from classical thermodynamics, which amounts to replacing $k_{\rm B} T$ with \mar{$\hbar\omega/[\exp(\hbar\beta\omega) -1]$} in Eq. (\ref{eq:equip}). 
Moreover, we should add a factor of 1/2 coming from vacuum fluctuations, resulting finally in
\begin{equation}
\mathbb{S}_{VV}(\omega )= 4 \hbar \omega \mathrm{Re}[Z(\omega )] \left(\frac{1}{e^{\beta\hbar\omega}-1}+\frac{1}{2}\right) = 2\hbar \omega \mathrm{Re}[Z(\omega )] \coth\frac{\beta \hbar \omega}{2}.
\end{equation}
Note that we still recover Eq. (\ref{eq:famous}) in the limit $\hbar \beta \omega \rightarrow 0$ since $\coth (\beta \hbar \omega /2)\rightarrow 2$. 

Comparing to our previous results, we find that this is the symmetrized single-sided spectral density
\begin{equation}
\mathbb{S}_{VV}(\omega )  = S_{VV}(\omega ) + S_{VV} (-\omega ).
\end{equation}

For ease of comparison with the full quantum results derived above, note that $n(\omega ) = \frac{1}{2}[\coth (\beta \hbar \omega/2) -1]$ and $n(\omega ) + 1 = \frac{1}{2}[\coth (\beta \hbar \omega/2) +1]$, with $n(\omega ) = 1/[\exp (\beta \hbar \omega )-1]$ being the number of photons. We also have the relation $n(-\omega ) = - n(\omega ) -1$. 
Also we note that the difference of these spectra is $S_{VV}(\omega ) - S_{VV}(-\omega ) = 2 \Re[Z(\omega)] \hbar\omega$, independent of temperature. This shows yet another connection between fluctuation and dissipation. For this result we use $\Re[Z(\omega )]= \Re[Z(-\omega )]$, which follows from Eq. (\ref{eq:zzstar}).

Finally, \mar{we would like to highlight a similarity between the phenomenological derivation of the Caldeira-Leggett model and Nyquist's derivation: the first Foster's form of a resistor is nothing but the normal mode decomposition of the electrical oscillations in the resistive circuit, as given in Eq.~\eqref{eqn:HamiltonianResistor}. Postulating that the resistor is in a bosonic thermal state means that, to each mode with frequency $\omega_j$, we assign an energy contribution of $\hbar\omega_j n(\omega_j)$ given by the Bose-Einstein distribution (see Eq.~\eqref{eqn:thermalStateSinglemode}). This is the quantum generalization of the standard equipartition principle. Analogously, Nyquist took into account all the modes of short-ended transmission line and assigned to each of them an energy contribution as in Eq.~\eqref{eq:equip}, according to the classical energy equipartition principle. The generalization of this to the quantum case requires employing the Bose-Einstein distribution, which again leads to the factor $\hbar\omega_j n(\omega_j)$ as above. We therefore see how both phenomenological derivations rely on similar arguments about mode decomposition and quantum energy distribution.} 

Both the Nyquist and the Caldeira-Leggett approaches are material-independent, that is to say, they do not consider the internal microscopic structure of the resistor, while being based on purely thermodynamic principles. The discussion of a microscopic model for the resistor and corresponding voltage fluctuations are the topic of the following section.

\subsection{Microscopic description of quantum fluctuations through a resistor}
In this section we  employ the Landauer-Büttiker method to describe the quantum fluctuations across a resistive element in a mesoscopic circuit. For simplicity, we  apply the method to a simple case, namely a DC voltage across a NIN junction, and then provide some more specialized references for the NIS junction or AC voltage.

\begin{figure}
\center
\includegraphics[scale=0.35]{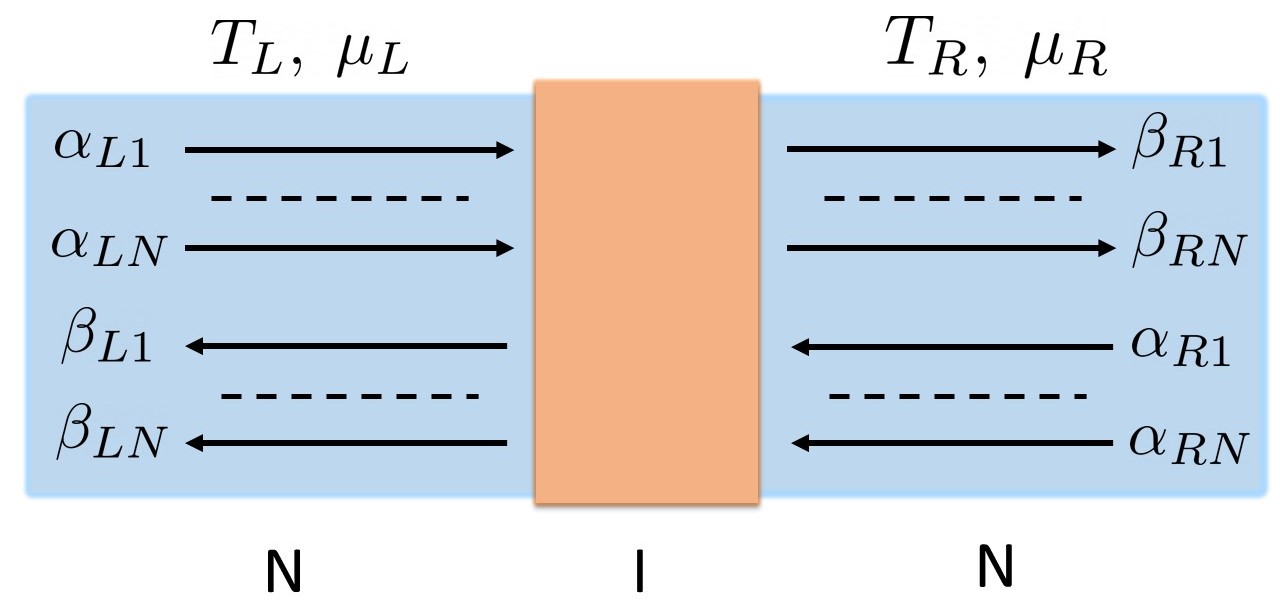}
\caption{Scattering channels in a NIN junction, where the left and the right leads are coupled to reservoirs with temperatures respectively $T_L$ and $T_R$ and chemical potentials respectively $\mu_L$ and $\mu_R$.}
\label{fig:scattering}
\end{figure}

Our aim is to write down the operator describing the current passage through a mesoscopic NIN junction and to compute its spectral density. We  follow the lines of the well-known review on the topic by Blanter and Büttiker \cite{Blanter2000}. We describe the passage of current as a scattering process across the insulator barrier, schematized as in Fig.~\ref{fig:scattering}: we assume that the electrons can run freely in the left (L) and right (R) leads, so that the total wavefunction (far from the barrier) can be written as a sum of free particles wavefunctions. We do not need to describe in details the scattering process inside and at the borders of the insulating barrier, while, for our purposes, it is sufficient to relate the output modes to the input modes through the scattering matrix. Specifically, using a second-quantization formalism, we call $\alpha_{L1},\ldots,\alpha_{LN}$ ($\alpha_{R1},\ldots,\alpha_{RN}$) the input modes in the left (right) lead, while $\beta_{L1},\ldots,\beta_{LN}$ ($\beta_{R1},\ldots,\beta_{RN}$) are the output modes, as shown in Fig.~\ref{fig:scattering}. $\alpha_{Ln,Rm}$ and $\beta_{Ln,Rm}$ are fermionic annihilation operators, and for simplicity we have taken an equal number of modes (``channels'') in the left and in the right lead. Let us insert them in two vectors as $\boldsymbol{\alpha}=(\alpha_{L1},\ldots,\alpha_{LN},\alpha_{R1},\ldots,\alpha_{RN})^T$, $\boldsymbol{\beta}=(\beta_{L1},\ldots,\beta_{LN},\beta_{R1},\ldots,\beta_{RN})^T$. Then, by looking at the characteristics of the scattering process, we can write the output modes as functions of the input modes:
\begin{equation}
\boldsymbol{\beta}=\mathbf{S}\boldsymbol{\alpha},
\end{equation}
where $\mathbf{S}$ is the \textit{scattering matrix}. We can write the latter as
\begin{equation}
\label{eqn:scattMat}
\mathbf{S}=\begin{pmatrix}
\mathbf{r}&\mathbf{t'}\\
\mathbf{t}&\mathbf{r'}\\
\end{pmatrix},
\end{equation}
where $\mathbf{r},\mathbf{r'}$ are block matrices describing the reflection coefficients in respectively the left and right lead, while $\mathbf{t}$ and $\mathbf{t'}$ describe the transmission ones.

The total wavefunction for the left lead for example can be written as:
\begin{equation}
\label{eqn:waveLeftlead}
\psi_{L}(\mathbf{x},t)=\frac{1}{\sqrt{2\pi}}\int dE \,e^{-\frac{i}{\hbar}E t}\sum_{n=1}^N\frac{\chi_{Ln}(x,y)}{\sqrt{2\pi\hbar v_{Ln}(E)}}\left(\alpha_{Ln}(E) e^{i k_{Ln}(E) z}+\beta_{Ln}(E) e^{-i k_{Ln}(E) z}\right),
\end{equation}
where $\chi_{Ln}(x,y)$ are transverse wavefunctions, $k_{Ln}(E)=\sqrt{2m(E-E_{Ln })}/\hbar$ is the free-particle momentum of the $n$th channel, and $v_{Ln}(E)=\hbar k_{Ln}(E)/m$ is the carrier velocity and $m$ the carrier mass. By making use of Eq.~\eqref{eqn:waveLeftlead}, we can write the current operator in the left lead as:
\begin{equation}
\label{eqn:currentOp}
I_{L}(\mathbf{x},t)=\frac{\hbar e}{2 i m}\int dx dy\,\left[ \psi_{L}^\dagger(\mathbf{x},t)\frac{\partial}{\partial z}\psi_{L}(\mathbf{x},t)-\frac{\partial}{\partial z}\psi_{L}^\dagger(\mathbf{x},t) \psi_{L}(\mathbf{x},t)\right].
\end{equation}
After some tedious algebra, and under the assumption that the carrier velocity varies slowly as a function of energy, we can rewrite Eq.~\eqref{eqn:currentOp} as \cite{Blanter2000}:
\begin{equation}
\label{eqn:currentFin}
I_{L}(t)=\frac{e}{2\pi\hbar}\sum_{\alpha,\beta=L,R}\sum_{n,m=1}^N\int dE dE' e^{\frac{i}{\hbar}(E-E')t} a_{\alpha n}^\dagger (E) A_{\alpha\beta}^{nm}(L;E,E') a_{\beta m}(E),
\end{equation}
with
\begin{equation}
\label{eqn:indexes}
A_{\alpha\beta}^{nm}(L;E,E')=\delta_{nm}\delta_{\alpha L}\delta_{\beta L}- \sum_k (\mathbf{S}^\dagger)_{L\alpha,nk}(E)(\mathbf{S})_{L\beta,km}(E'),
\end{equation}
where $(\mathbf{S}^\dagger)_{L\alpha,nk}(E)$ is the coefficient describing the dependence of $b_{Ln}(E)$ on $a_{\alpha k}(E)$.

Starting from Eq.~\eqref{eqn:currentFin}, we can calculate all the non-equilibrium features of the current across the mesoscopic NIN junction. In particular, let us assume that the left lead is connected to a reservoir with temperature $T_L$ and chemical potential $\mu_L$ as shown in Fig.~\ref{fig:scattering} (and correspondingly for the right lead). Then, we can compute the spectral density of the operator $I_{L}(t)$ in Eq.~\eqref{eqn:currentFin}, defined as:
\begin{equation}
\label{eqn:spectIdef}
S_{I_LI_L}(\omega)=\int_{-\infty}^\infty dt\, e^{i\omega t} \langle I_{L}(t)I_{L}(0)\rangle.
\end{equation}
For the zero frequency case, current conservation immediately implies $S_{I_RI_R}(0)=S_{I_LI_L}(0)$, therefore without losing generality we can focus on the latter. 
After some long but trivial manipulations of Eq.~\eqref{eqn:currentFin} in the definition of Eq.~\eqref{eqn:spectIdef}, it can be shown \cite{Blanter2000} that, for zero frequency and zero applied voltage, the spectral density has the following form:
\begin{equation}
\label{eqn:spectDenI}
\begin{split}
S_{I_LI_L}(0)=\frac{e^2}{\pi\hbar} \sum_{n=1}^N\int dE\left[t_n(E)(f_L(E)(1-f_L(E))+f_R(E)(1-f_R(E))\right],
\end{split}
\end{equation}
where $t_n(E)$ are the eigenvalues of the matrix $\mathbf{t}^\dagger\mathbf{t}$, while $f_L(E)$ is the Fermi-Dirac distribution in the left lead at energy $E$. In the common case where $t_n(E)$ varies slowly as a function of energy, we can replace it with its value at the Fermi energy, $t_n(E_F)$. Integrating the Fermi-Dirac distributions in the equilibrium scenario ($T_L=T_R$), we obtain:
\begin{equation}
\label{eqn:spectDenFinI}
\begin{split}
S_{I_LI_L}(0)=\frac{2e^2k_BT}{\pi\hbar} \sum_{n=1}^N t_n(E_F)=2k_B T G,
\end{split}
\end{equation}
where $G=\frac{e^2}{\pi\hbar} \sum_{n=1}^N t_n(E_F)$ is the conductance according to Landauer's formula \cite{Nazarov2009}. Now we can compare Eq.~\eqref{eqn:spectDenFinI} with Eq.~\eqref{eqn:spectrumResistor} (the latter in the low-frequency regime): they both display the Johnson-Nyquist thermal noise at equilibrium (the former for  current while the latter for voltage), although they have been obtained through completely different approaches. The spectrum of Eq.~\eqref{eqn:spectrumResistor} has been derived following a phenomenological model, based on the impedance behavior of the resistor in the lumped element model of the circuit, and makes use of an effective temperature of the resistor. On the contrary, Eq.~\eqref{eqn:spectDenFinI} has been derived through a microscopic physical model describing the scattering of charge carriers in a NIN junction, and the temperature appearing in the noise formula is the temperature of the left and right leads of the junction. Similar results can be found in the case of non-zero frequency spectral density \cite{Buttiker1992} or NIS junctions \cite{Khlus1987,DeJong1994}, leading to Eq.~\eqref{eqn:spectrumResistor}. We can therefore conclude that the microscopic model discussed in this section provides us with a good intuitive physical motivation of the Caldeira-Leggett model used throughout the paper. 

\mar{In general terms, the Caldeira-Leggett model fails to reliably describe the circuit dynamics when the so-called ``lumped element model'' breaks down, i.e., when the microscopic behavior of the carriers in the circuit cannot be described anymore by well-separated capacitive, inductive, and resistive elements. This approximation is widely employed in the study of circuit quantum electrodynamics \cite{Blais2020}. A detailed discussion about the validity of the lumped element model in electrical engineering is provided in Ref.~\cite{Agarwal2005}, while some related comments for superconducting circuits can be found, for instance, in Refs.~\cite{Girvin2014,Rasmussen2021}.} 

\section{Capacitive coupling in circuit QED}
\label{sec:capCoupling}
In order to understand how to couple superconducting qubits with resistors, which is the subject of Sec.~\ref{sec:coupling}, we first have to get familiar with the capacitive coupling. Indeed, although several ways of coupling elements in circuit QED are available \cite{Wendin2007a}, the capacitive coupling is the natural one when we want to connect transmon qubits and resistors. 

\begin{figure}
\center
\includegraphics[scale=0.55]{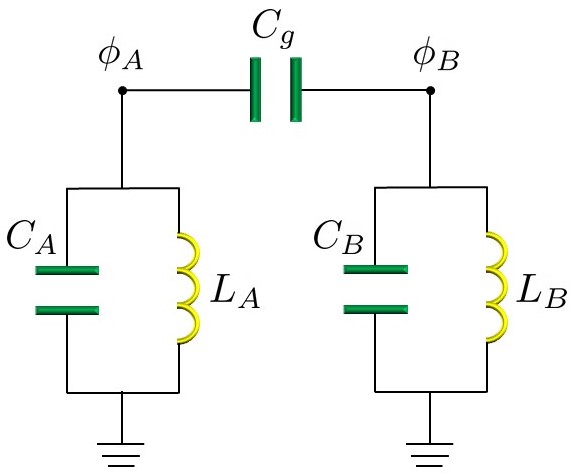}
\caption{Two parallel LC circuits coupled via a capacitor with capacitance $C_g$.}
\label{fig:capacitiveCoupling}
\end{figure}

First of all, let us discuss how the capacitive coupling can be realized between two LC circuits \cite{Girvin2014}. The coupling scheme is depicted in Fig.~\ref{fig:capacitiveCoupling}: two LC circuits, named $A$ and $B$, are connected via a capacitor with capacitance $C_g$. Following the standard quantization procedure for quantum circuits, we first of all have to write the Lagrangian of the system and then find the Hamiltonian through Legendre transformation. We refer the reader to Appendix~\ref{sec:quantization} for a detailed presentation of this procedure. The system Lagrangian reads:
\begin{equation}
\label{eqn:LagrangianCouplingLC}
\begin{split}
\mathcal{L}&=\frac{C_A}{2}\dot{\phi}_A^2-V(\phi_A)+\frac{C_B}{2}\dot{\phi}_B^2-V(\phi_B)+\frac{C_g}{2}\left(\dot{\phi}_A-\dot{\phi}_B\right)^2\\
&=\frac{1}{2}\dot{\vecP}^T C\dot{\vecP}-\sum_{i=A,B}V(\phi_i),
\end{split}
\end{equation}
where we have introduced the \textit{flux vector} $\vecP=(\phi_A,\phi_B)^T$ and the \textit{capacitance matrix} defined as
\begin{equation}
\label{eqn:capacitanceMatrix}
C=\begin{pmatrix}
C_A+C_g&-C_g\\
-C_g&C_B+C_g
\end{pmatrix}.
\end{equation}
Note that, for convenience, we express the potential energies of the inductors as $V(\phi_i)=\frac{1}{2L_i}\phi_i^2$.

After having obtained the momenta $Q_i=\sum_{j=A,B}C_{ij}\dot{\phi}_j$, collected in the vector $\vecQ=(Q_A,Q_B)^T$, we write the Hamiltonian of the system as
\begin{equation}
\label{eqn:HamiltonianCouplingLC}
\mathcal{H}=\frac{1}{2}\vecQ^T C^{-1} \vecQ+\sum_{i=A,B}V(\phi_i),
\end{equation}
and we can proceed with the quantization of the appropriate conjugate variables: $\{\phi_j,Q_k\}=\delta_{jk}\rightarrow [\phi_j,Q_k]=i\hbar\delta_{jk}$.
Note that, unlike in the case of a simple LC circuit discussed in Appendix~\ref{sec:quantization}, each momentum $Q_j$ is not proportional to the voltage across the $j$th LC circuit $\dot{\phi}_j$, but is a more involved function of both the voltage differences. In particular, $\{\phi_j,\dot{\phi}_k\}=C^{-1}_{kj}$, and $C^{-1}$ reads:
\begin{equation}
C^{-1}=\frac{1}{C_AC_B+C_AC_g+C_BC_g}\begin{pmatrix}
C_B+C_g&C_g\\
C_g&C_A+C_g
\end{pmatrix}.
\end{equation}
We observe that the voltage across each LC circuit can still be considered as the conjugate variable with respect to the node flux, up to a constant with the units of the inverse of capacitance, if the coupling capacitance is low, i.e. under the weak coupling condition $C_g\ll C_A,C_B$. In this limit, we assume that $C_A$ and $C_B$ are of the same order, i.e. $O(C_B/C_A)=O(1)$; therefore, without losing generality, we choose to employ the perturbation order $O(C_g/C_A)$. For instance, the weak coupling condition leads to $\{\phi_j,\dot{\phi}_k\}=C_{jj}^{-1}\delta_{kj}+O(C_g/C_A)$, equivalent to $Q_j=C_{jj}\dot{\phi}_j+O(C_g/C_A)$. Note that the above description is valid for any potential energy $V(\phi_j)$.

The off-diagonal elements in $C^{-1}$ represent couplings between the quantized modes of the system. For instance, with $V(\phi_j)=\frac{1}{2L_i}\phi_i^2$, we can see that the off-diagonal elements correspond to a coupling between the $p$ quadratures\footnote{The quadrature $p$ corresponds to the momentum $Q$ in the notation of circuit QED. We are using the notation of Ref. \cite{MandelWolf} for the quantum electromagnetic field, which mimics the usual notation for the quantum harmonic oscillator. This will be useful when discussing the coupling to a thermal bath.} of two quantum harmonic oscillators, in the form $p_Ap_B$.

\begin{figure}[H]
\center
\includegraphics[scale=0.55]{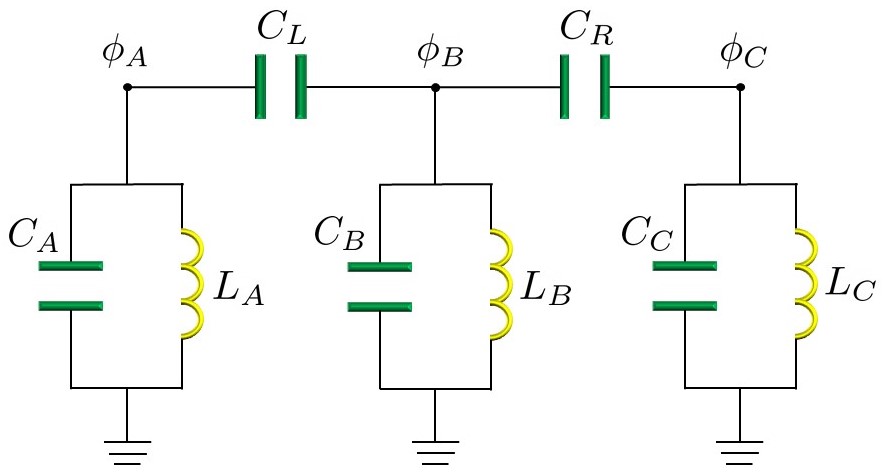}
\caption{Three parallel LC circuits in a line, with two coupling capacitors $C_L$ and $C_R$. Although from a ``diagrammatic point of view'' we may say that the LC circuits are connected exclusively to the first neighbors in the line, in the Hamiltonian this is true only in the weak coupling limit, while in general a strong coupling between the three modes of the circuit emerges.}
\label{fig:3LCcircuits}
\end{figure}

The description relying on the flux vector and capacitance matrix is valid for a general number of capacitively coupled LC circuits. The reader can verify that, for instance, if we deal with three LC circuits coupled capacitively in a line as in Fig.~\ref{fig:3LCcircuits}, the Hamiltonian of the system includes coupling between all the three modes, but in the weak coupling limit $C_L,C_R\ll C_A,C_B,C_C$ we find only first-neighbor coupling of the order of ${O}(C_{L,R}/C_{A,B,C})$, while the second-neighbor coupling is of the order of ${O}((C_{L,R}/C_{A,B,C})^2)$, and therefore it can be neglected. More generally, we can summarize the action of the capacitive coupling as: when we connect LC circuits in a line through capacitors, the voltage fluctuations across each circuit are coupled only to the first neighbours. This is reflected in the matrix $C$ appearing in the Langrangian. The conjugate variables, however, are not the voltage fluctuations but the momenta, which are all coupled between each other as described by the matrix $C^{-1}$. Therefore, in general, the Hamiltonian displays a long-range coupling between the modes of the system. In the weak coupling limit, i.e. when all the coupling capacitors are far weaker than the capacitors of each LC circuit, the long-range interaction is broken and the Hamiltonian describes a first-neighbour coupling as well.

Trivially, the statement above holds not only for LC circuit, but also for superconducting qubits, i.e. when the inductive energy $V(\phi_j)$ has a contribution given by a Josephson junction. In the next section, we will observe a very similar behaviour when discussing the coupling between qubits and resistors.

\section{Coupling qubits to resistors as thermal baths}
\label{sec:coupling}

In this section we  introduce some relevant examples of resistors acting as thermal baths coupled to superconducting qubits. We  focus on the most basic scenarios with one or two qubits, in particular addressing the case of one qubit coupled to a bath, of two qubits coupled to a common bath, with and without a direct coupling between them, and finally of two coupled qubits interacting with separate baths. 

For each case, we  present the Hamiltonian of the system and, when possible, the associated master equation describing the dynamics of the qubits. We do not provide here the derivation of each Hamiltonian, which is quite involved and can be found in Appendix~\ref{sec:derivation}. Note that all the cases discussed in this section assume weak coupling between qubits and baths, and for this reason we focus here on \textit{Markovian} master equations \cite{BreuerPetruccione}. \mar{As we will see, the weak coupling limit is controlled by the ratio between the qubit-resistor coupling capacitance and the internal qubit capacitance.} The treatment of the strong coupling limit under the Caldeira-Leggett model, \mar{emerging when the qubit-resistor coupling capacitance is of the same order as the internal qubit capacitance,} requires a numerical derivation of both the circuit Hamiltonian and of the qubit dynamics - see for instance the discussion in Ref.~\cite{Pekola2020}. The experimental realization of a strong coupling is possible as well, being simply related to the value of the coupling capacitor. In this regime, however, the final form of the Hamiltonian is more involved, and sometimes it may not even be possible to express it in a closed form through the Caldeira-Leggett method, as we show in Appendix~\ref{sec:derivation}. \mar{The open system dynamics in the presence of strong coupling with the bath typically displays non-Markovian dynamics \cite{PhysRevB.103.214308}, revivals, and has crucial consequences also for the study of quantum thermodynamics \cite{PhysRevE.86.061132,PhysRevB.90.075421,PhysRevLett.124.160601}. This regime has been numerically tested in the case where the open system is a single superconducting qubit, for instance, in Refs.~\cite{PhysRevResearch.1.013004,Pekola2020}.}

Finally, note that, throughout the paper, we  represent a superconducting qubit as a capacitor in parallel with an inductive ``potential energy'' $V$. Not writing the explicit form of $V$ is quite convenient, because, in this way, we can describe different circuit elements by varying $V$, such as a qubit or a LC circuit, and at the same time the method of deriving the interaction Hamiltonian does not depend on $V$. More details about it can be found in Appendix~\ref{sec:quantization} and in Ref.~\cite{devoret1995}. For each scenario, we  provide the Hamiltonian keeping the abstract notation with the potential energy $V$, and then we  rewrite it in the specific scenario with superconducting transmon qubits. Therefore, our results may be trivially extended to the case of harmonic oscillators coupled to thermal baths, by replacing the transmon qubits with LC circuits. 
 
\subsection{Single qubit}
\label{sec:singleQ}

\begin{figure}
\center
\includegraphics[scale=0.55]{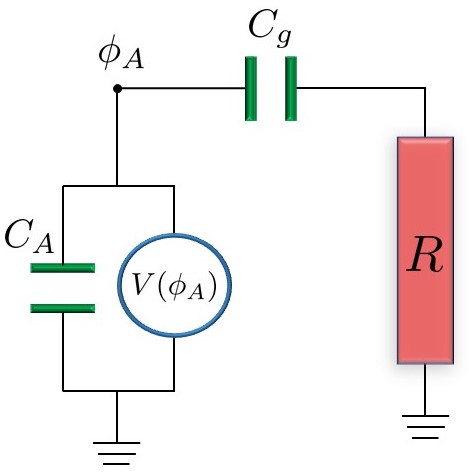}
\caption{A single qubit $A$ coupled to a resistor. The resistor is described according to the Foster's first form, and the node fluxes of the internal degrees of freedom (in the present picture not shown) are the ones depicted in Fig.~\ref{fig:firstFoster}.}
\label{fig:oneQubit}
\end{figure}

Fig.~\ref{fig:oneQubit} depicts the circuit representing a single qubit, which we name as A, capacitively coupled to a resistor with resistance $R$. The capacitance of the coupling capacitor is $C_g$. The coordinates of the systems are the flux $\phi_A$ associated to qubit A as in Fig.~\ref{fig:oneQubit}, and the ``fictional'' infinite non-interacting internal variables of the resistor. We term as $p_A$ the momentum corresponding to $\phi_A$. In the Hamiltonian, we describe the internal variables of the resistor as modes labelled by $\alpha$ and created by the operator $a_\alpha^\dagger$. The details of the calculation to obtain the final Hamiltonian can be found in Appendix~\ref{sec:appendixSingle}.

We now consider the weak coupling limit, which, eluding some subtleties discussed in Appendix~\ref{sec:weakcoupling}, reads $C_g\ll C_A$. Neglecting all the terms of the order of $O((C_g/C_A)^2)$, the circuit Hamiltonian is given by:
\begin{equation}
\label{eqn:weakHamSingleQtext}
\begin{split}
\mathcal{H}=&\frac{p_A^2}{2C_{\Sigma_A}}+V(\phi_A)+\sum_{\alpha=1}^\infty \hbar\omega_\alpha a_\alpha^\dagger a_\alpha+i\frac{C_g}{C_{\Sigma_A}}\sum_{\alpha=1}^\infty  \sqrt{\frac{\hbar\omega_\alpha \Re[Z(\omega_\alpha)]\Delta\omega}{\pi}}p_A (a_\alpha^\dagger-a_\alpha) ,\\
\end{split}
\end{equation}
where $C_{\Sigma_A}=C_A+C_g$, $\Delta\omega$ is the infinitesimal frequency gap introduced in the Caldeira-Leggett model of a resistor in Eq.~\eqref{eqn:valueDevoretComega}, which will be absorbed when computing the bath spectral density, and $\omega_\alpha=\alpha\Delta\omega$ are the frequencies of the internal modes of the resistor, forming a continuum in the limit $\Delta\omega\rightarrow 0$. If we choose as $V(\phi_A)$ the inductive energy of a transmon qubit \cite{Koch2007a} with frequency $\omega_A$, we can rewrite Eq.~\eqref{eqn:weakHamSingleQtext} as:
\begin{equation}
\label{eqn:weakHamSingleQqubit}
\begin{split}
\mathcal{H}=&\frac{\hbar\omega_A}{2}\sigma_A^z+\sum_k \hbar\omega_k a_k^\dagger a_k+i\lambda_A\frac{C_g}{C_{\Sigma_A}}\sum_k\sqrt{\frac{R\hbar\omega_k\omega_C^2\Delta\omega}{\pi(\omega_C^2+\omega_k^2)}}\sigma_A^y(a_k^\dagger-a_k),
\end{split}
\end{equation}
where we have replaced the spectrum of the resistor with Eq.~\eqref{eqn:impedanceOhmic} and we have introduced a summation over the ``dense'' index $k$, as usually done in quantum optics. $\lambda_A$ is a constant with the units of charge, depending on the features of the transmon qubit\footnote{To be more precise, supposing that the transmon qubit Hamiltonian can be approximated to that of a harmonic oscillator with plasma frequency $\omega_A=\sqrt{8E_CE_J}/\hbar$, where $E_C$ and $E_J$ are respectively the charging and Josephson energy \cite{Koch2007a}, we have $\lambda_A\approx\sqrt{\hbar\omega_A C_{\Sigma_{A}}/2}$.}, which allows us to pass from $p_A$ to $\sigma_A^y$ through $p_A\approx \lambda_A\sigma_A^y$ \cite{Koch2007a}. We observe that Eq.~\eqref{eqn:weakHamSingleQqubit} describes the standard dissipative spin-boson model with an Ohmic spectral density of the bath \cite{BreuerPetruccione}, defined as:
\begin{equation}
\label{eqn:spectralDensitySingleQ}
\begin{split}
J(\omega)&=\gamma\sum_{k} \frac{\omega_k\omega_C^2\Delta\omega}{\pi(\omega_C^2+\omega_k^2)}\delta(\omega-\omega_k)=\gamma\int_0^\infty d\omega_k \frac{\omega_k\omega_C^2}{\pi(\omega_C^2+\omega_k^2)}\delta(\omega-\omega_k)=\gamma\frac{\omega\omega_C^2}{\pi(\omega_C^2+\omega^2)},
\end{split}
\end{equation}
where we have used the infinitesimal frequency gap $\Delta\omega$ to transform the summation into an integral, and we have introduced:
\begin{equation}
    \gamma= \frac{R \lambda_A^2}{\hbar} \frac{C_g^2}{C_{{\Sigma}_A}^2}.
\end{equation}

 If the resistor is in a thermal state with temperature $T$, in the weak coupling limit the master equation describing the state of the qubit at time $t$, $\rho_A(t)$, is \cite{BreuerPetruccione}:
\begin{equation}
\label{eqn:masterEqSingleQ}
\begin{split}
\frac{d\rho_A(t)}{dt}=&-i[\omega_A/2\sigma_A^z+H_{LS}/\hbar,\rho_A(t)]+\mar{\Gamma^{\downarrow}(\beta )} \underbrace{\left(\sigma_A^-\rho_A(t)\sigma_A^+-\frac{1}{2}\{\sigma_A^+\sigma_A^-,\rho_A(t)\}\right)}_{\textnormal{emission}}\\
&+\mar{\Gamma^{\uparrow}(\beta )} \underbrace{\left(\sigma_A^+\rho_A(t)\sigma_A^--\frac{1}{2}\{\sigma_A^-\sigma_A^+,\rho_A(t)\}\right)}_{\textnormal{absorption}}.
\end{split}
\end{equation}
$H_{LS}$ is the Lamb-Shift Hamiltonian, which only induces a frequency-shift \cite{BreuerPetruccione}. \mar{$\Gamma_\downarrow (\beta )$ and $\Gamma_\uparrow (\beta )$} are respectively the emission and absorption rates, which depend on the bath temperature. By looking at the qubit-bath interaction Hamiltonian in Eq.~\eqref{eqn:weakHamSingleQqubit}, we observe that it can be rewritten as $\mathcal{H}_I=\lambda_A C_g/C_{\Sigma_A}\sigma_A^yV$, where $V$ is the operator describing the overall voltage across the resistor, i.e. the one whose correlation function we have computed in Eq.~\eqref{eqn:correlationFunctionResistor}. The form of the correlation function, and therefore of the spectral density, does not change in the weak coupling limit \cite{Jones2013}. Hence, the dissipation induced on the qubit is driven by the fluctuations of the voltage difference (see for instance the derivation in Refs.~\cite{Jones2013,Tuorila2017,Ronzani2018}), and we can think of the qubit as a quantum spectrum analyser \cite{Clerk2010}; the decaying rates are readily obtained:
\begin{equation}
\label{eqn:decayRates}
\begin{split}
&\mar{\Gamma^{\downarrow}(\beta )}=\left(\frac{\lambda_A C_g}{\hbar C_{\Sigma_A}}\right)^2 S_{VV}(\omega_A)=\pi J(\omega_A)\left(\coth\frac{\beta\hbar\omega_A}{2}+1\right),\\
&\mar{\Gamma^{\uparrow}(\beta )}=\left(\frac{\lambda_A C_g}{\hbar C_{\Sigma_A}}\right)^2 S_{VV}(-\omega_A)=\pi J(\omega_A)\left(\coth\frac{\beta\hbar\omega_A}{2}-1\right),\\
\end{split}
\end{equation}
where $S_{VV}(\omega)$ is the \mar{temperature-dependent} spectral density in Eq.~\eqref{eqn:spectrumResistor}. \mar{Note that $\Gamma^{\downarrow}(\beta )= \Gamma(n (\omega ) +1)$ and $\Gamma^{\uparrow}(\beta )= \Gamma n (\omega )$, where $\Gamma = \pi J(\omega_{\rm A})$ is the zero-temperature decay rate, $\Gamma=\Gamma^{\downarrow}(\beta = \infty)$.
We also have $\Gamma^{\uparrow}(\beta ) = \exp(-\beta \hbar \omega)\Gamma^{\downarrow}(\beta )$, reflecting the detailed balance principle from thermodynamics.}

\subsubsection{Adding an LC filter}
\begin{figure}
\center
\includegraphics[scale=0.55]{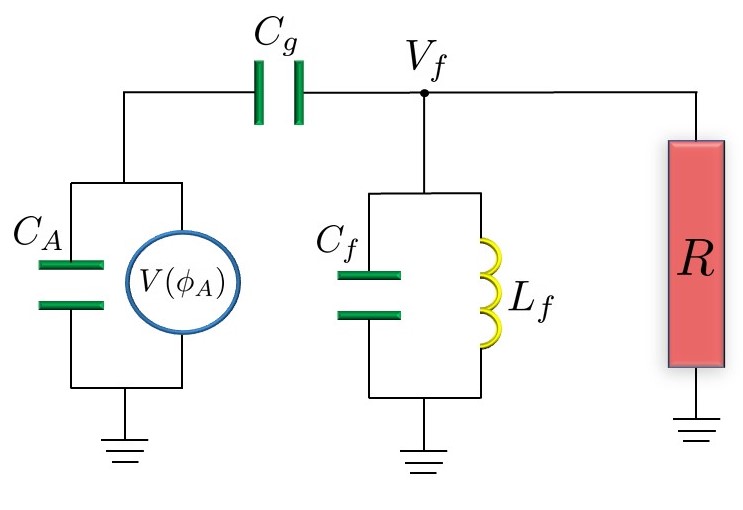}
\caption{Superconducting circuit as in Fig.~\ref{fig:oneQubit}, but with an additional LC circuit between the transmon qubit and the resistor, playing the role of a spectral filter.}
\label{fig:filter}
\end{figure}
The spectral density in Eq.~\eqref{eqn:spectralDensitySingleQ} can be shaped by inserting a LC circuit between qubit and resistor as shown in Fig.~\ref{fig:filter}. The LC circuit, with frequency $\omega_f=1/\sqrt{C_f L_f}$, capacitance $C_f$ and inductance $L_f$, is directly coupled to the transmon qubit through the capacitor $C_g$. Once again, we assume the weak coupling limit, i.e. $C_g\ll C_A,C_f$, where $C_A$ is the total capacitance of the qubit. Following the discussion for the previous case, under this assumption the qubit operator $\sigma_A^y$ gets coupled to the operator describing the voltage $V_f$ at the LC circuit node (see Fig.~\ref{fig:filter}) in the final Hamiltonian. The bath spectral density of this dissipative coupling can be obtained, according to Eqs.~\eqref{eqn:weakHamSingleQtext}, ~\eqref{eqn:spectralDensitySingleQ} and~\eqref{eqn:decayRates}, by evaluating the autocorrelation function of the voltage fluctuations $\langle V_f(t)V_f(0)\rangle$ and then the corresponding spectral density $S_{V_fV_f}(\omega)$. The latter can be derived by analysing the circuit in Fig.~\ref{fig:filter} where, following the convention in electrical engineering, the noisy resistor can be decomposed as a noiseless resistor with resistance $R$ in series with a voltage noise source with suitable spectral density. In the weak coupling limit, the current passing through the small capacitor $C_g$ is negligible, therefore we can forget about the transmon qubit and focus on the circuit on the right-hand side of $C_g$; then, we can find the value of $V_f(t)$ as a function of $V(t)$, where the latter is the voltage produced by the noise source, whose spectral density Eq.~\eqref{eqn:spectrumResistor} is known. If $Z_f(\omega)$ is the impedance of the parallel LC circuit (see Eq.~\eqref{eqn:impedanceLC}), then current conservation under the weak coupling limit yields:
\begin{equation}
    V_f(\omega)=\frac{Z_f(\omega)}{Z_f(\omega)+R} V(\omega)=h_{f}(\omega)V(\omega),
\end{equation}
where $h_{f}(\omega)$ is the transfer function. Therefore, $S_{V_fV_f}(\omega)=\abs{h_{f}(\omega)}^2 S_{VV}(\omega)$, and the open dynamics of the transmon qubit is described, once again, by the spin-boson model, with the prescription that the bath spectral density is modified as $J(\omega)\rightarrow\abs{h_{f}(\omega)}^2 J(\omega)$. Then, the decay rates can be found through the same formula Eq.~\eqref{eqn:decayRates}. A simple calculation gives
\begin{equation}
    \label{eqn:transferFunction}
    \abs{h_{f}(\omega)}^2=\frac{\omega^2}{\omega^2+R^2C_f^2(\omega_f^2-\omega^ 2)^2},
\end{equation}
where $\abs{h_{f}(\omega)}^2$ is a Lorentzian distribution  resulting in band-pass filtering with bandwidth $\Delta\omega=1/RC_f$ and quality factor $Q_f = \omega_{f}/\Delta\omega = R\sqrt{C_{f}/L_{f}}$.
Several other schemes of RLC filters with different transfer functions can be engineered in order to obtain the desired spectral density \cite{dimopoulos2011analog}. \mar{They can also be studied mathematically by means of the Vernon transform \cite{Aurell2021}.}

This filtering scheme can be really useful, for instance, to avoid leakages to higher transmon levels, see e.g. the similar strategies for the theoretical proposals in Refs.~\cite{Jones2013,Tuorila2017} and for the experiments in Refs.~\cite{Ronzani2018,Senior2019}. Several other schemes that use filters are discussed in Section V.

\subsection{Two qubits, common bath}
\label{sec:twoQcommon}
\subsubsection{No direct coupling}
\begin{figure}[t]
\center
\includegraphics[scale=0.55]{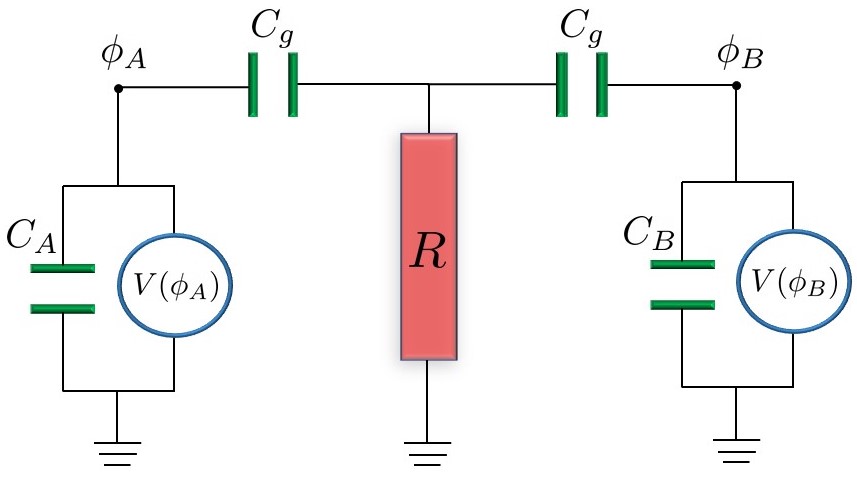}
\caption{Two qubits $A$ and $B$ coupled to a common resistor. The resistor is described according to the Foster's first form, and the node fluxes of the internal degrees of freedom (in the present picture not shown) are the ones depicted in Fig.~\ref{fig:firstFoster}.}
\label{fig:twoQ}
\end{figure}
Two superconducting qubits coupled to a common resistor through two distinct capacitive couplings, and without a direct coupling between them, are shown in Fig.~\ref{fig:twoQ}. For simplicity, we have chosen the same value $C_g$ for both the coupling capacitors, but other choices are possible, as discussed in Appendix~\ref{sec:appendixTwoQComm}. The system coordinates are the fluxes $\phi_A$ and $\phi_B$ associated respectively to qubit A and qubit B as in Fig.~\ref{fig:twoQ}, and, as always, the ``fictional'' infinite non-interacting internal variables of the resistor. We term $p_A$ and $p_B$ the momenta corresponding respectively to $\phi_A$ and $\phi_B$. In the Hamiltonian, we describe the internal variables of the resistor as modes labelled by $\alpha$ and created by the operator $a_\alpha^\dagger$. In Appendix~\ref{sec:appendixTwoQComm} we extensively discuss how to obtain the correct momenta, the modes of the resistor and the final Hamiltonian of the system. In the weak coupling limit, expressed by the condition $C_g\ll C_A,C_B$, such Hamiltonian reads:
\begin{equation}
\label{eqn:weakCouplTwoQtext}
\begin{split}
\mathcal{H}=&\frac{1}{2C_{\Sigma_A}}p_A^2+V(\phi_A)+\frac{1}{2C_{\Sigma_B}}p_B^2+V(\phi_B)+\sum_{\alpha=1}^\infty \hbar\omega_\alpha a_\alpha^\dagger a_\alpha\\
&+iC_g\sum_{\alpha=1}^\infty\sqrt{\frac{\Re[Z(\omega_\alpha)]\hbar\omega_\alpha\Delta\omega}{\pi}}\left(\frac{p_A}{C_{\Sigma_A}}+\frac{p_B}{C_{\Sigma_B}}\right)(a_\alpha^\dagger-a_\alpha),
\end{split}
\end{equation}
where $C_{\Sigma_j}=C_j+C_g$, and $\omega_\alpha=\alpha\Delta\omega$ according to Eq.~\eqref{eqn:valueDevoretComega}. The weight of each mode in the interaction Hamiltonian leads to an Ohmic spectral density as in Eq.~\eqref{eqn:spectralDensitySingleQ}.

As for the single qubit case, we express the circuit Hamiltonian by choosing as $V(\phi_A)$ and $V(\phi_B)$ the inductive energy of two transmon qubits \cite{Koch2007a} of frequency $\omega_A$ and $\omega_B$:
\begin{equation}
\label{eqn:weakHamTwoQqubit}
\begin{split}
\mathcal{H}=&\frac{\hbar\omega_A}{2}\sigma_A^z+\frac{\hbar\omega_B}{2}\sigma_B^z+\sum_{k} \hbar\omega_k a_k^\dagger a_k\\
&+iC_g\sum_{k}\sqrt{\frac{\Re[Z(\omega_k)]\hbar\omega_k\Delta\omega}{\pi}}\left(\frac{\lambda_A\sigma_A^y}{C_{\Sigma_A}}+\frac{\lambda_B\sigma_B^y}{C_{\Sigma_B}}\right)(a_k^\dagger-a_k),
\end{split}
\end{equation}
where $\lambda_A$ and $\lambda_B$ are constants with the units of charge, depending on the intrinsic values of the energies of the transmon qubits\footnote{According to the discussion for the single qubit case, we have $\lambda_j=\sqrt{\hbar\omega_j C_{\Sigma_j}/2}$, where $\omega_j$ is the plasma frequency of each transmon qubit \cite{Koch2007a}.}, given by the substitution of $p_j$ by $\sigma_j^y$, through $p_j\approx \lambda_j\sigma_j^y$. Eq.~\eqref{eqn:weakHamTwoQqubit} describes the standard Hamiltonian of two qubits interacting through a common bath, which is at the basis of several collective quantum phenomena \cite{Cattaneo2020}, such as superradiance \cite{Dicke1954}, entanglement generation \cite{Benatti2003} and quantum synchronization \cite{Giorgi2013}.

The master equation driving the evolution of the two-qubit state $\rho_{AB}$ is given by \cite{Cattaneo2019}:
\begin{equation}
\label{eqn:masterEqNoDir}
\begin{split}
\frac{d}{dt}\rho_{AB}(t)=&-\frac{i}{\hbar}[H_S+H_{LS},\rho_{AB}(t)]+\underbrace{\sum_{j,k=A,B}\mar{\Gamma_{jk}^\downarrow(\beta )}\left(\sigma_j^-\rho_{AB}(t)\sigma_k^+-\frac
{1}{2}\{\sigma_k^+\sigma_j^-,\rho_{AB}(t)\}\right)}_{\textnormal{emission}}\\
&+\underbrace{\sum_{j,k=A,B}{\mar{\Gamma}^\uparrow_{jk}(\beta )}
\left(\sigma_j^+\rho_{AB}(t)\sigma_k^- 
-\frac{1}{2}\{\sigma_k^-\sigma_j^+,\rho_{AB}(t)\}\right),}_{\textnormal{absorption}}
\\
\end{split}
\end{equation}
where $H_S$ is the free Hamiltonian of the two qubits, $H_{LS}$ is the Lamb-Shift Hamiltonian \cite{BreuerPetruccione} and \mar{$\Gamma_{jk}^\downarrow (\beta )$, ${\Gamma}^\uparrow_{jk}(\beta )$} are the coefficients of the master equation, which depend on the spectral density Eq.~\eqref{eqn:spectralDensitySingleQ}, on the qubit frequencies and on temperature of the bath, and they fulfil the condition \mar{$\Gamma_{jk}^{\downarrow\uparrow}(\beta )=\left[\Gamma_{kj}^{\downarrow\uparrow}(\beta )\right]^{*}$}. We refer the interested reader to Ref.~\cite{Cattaneo2019} for their explicit form.

Note that the behaviour of the capacitive coupling between qubits and resistor is analogous to the one observed in Sec.~\ref{sec:capCoupling}: in the weak coupling limit $C_g\ll C_A,C_B$ we do not have a direct long-range interaction between the two qubits, while, as proven in Appendix~\ref{sec:appendixTwoQComm}, for strong coupling this is not true. In many situations, we are actually interested in having a direct qubit-qubit coupling in the Hamiltonian. In the next section we show how to obtain it by inserting an additional coupling capacitor. 

\subsubsection{Direct coupling}
\begin{figure}
\center
\includegraphics[scale=0.55]{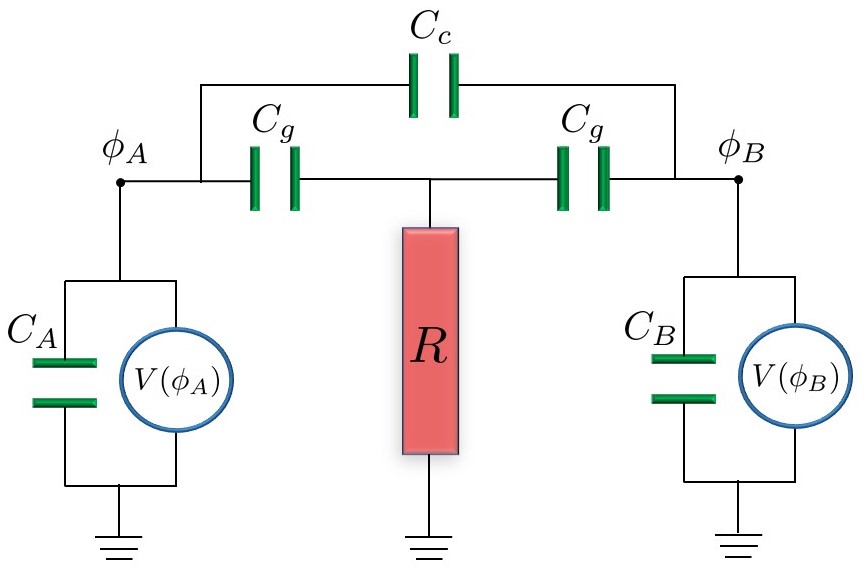}
\caption{Two qubits $A$ and $B$ coupled to a common resistor, with an additional coupling capacitor $C_c$ directly connecting them. The latter is necessary to induce a direct coupling $\sigma_A^y\sigma_B^y$ between the qubits in the weak coupling limit. The resistor is described according to the Foster's first form, and the node fluxes of the internal degrees of freedom (in the present picture not shown) are the ones depicted in Fig.~\ref{fig:firstFoster}.}
\label{fig:twoQdirect}
\end{figure}
Let us consider the circuit analysed in the previous section and add an additional (weak or strong) coupling capacitor $C_c$ as in Fig.~\ref{fig:twoQdirect}. In the same weak coupling limit, i.e. assuming $C_g\ll C_A,C_B$, it can be proven that such capacitor induces a direct qubit-qubit coupling (see Appendix~\ref{sec:appendixTwoQComm} for details). Using the same notation as in the previous section, the Hamiltonian, keeping the general form of the inductive potentials $V(\phi_j)$, reads:
\begin{equation}
\label{eqn:weakCouplTwoQDir}
\begin{split}
\mathcal{H}=&\frac{\mathsf{S}^{-1}_{11}}{2}p_A^2+V(\phi_A)+\frac{\mathsf{S}^{-1}_{22}}{2}p_B^2+V(\phi_B)+\mathsf{S}^{-1}_{12}p_Ap_B+\sum_{\alpha=1}^\infty \hbar\omega_\alpha a_\alpha^\dagger a_\alpha\\
&+iC_g\sum_{\alpha=1}^\infty\sqrt{\frac{\Re[Z(\omega_\alpha)]\hbar\omega_\alpha\Delta\omega}{\pi}}\left((\mathsf{S}^{-1}_{11}+\mathsf{S}^{-1}_{12})p_A+(\mathsf{S}^{-1}_{21}+\mathsf{S}^{-1}_{22})p_B\right)(a_\alpha^\dagger-a_\alpha),
\end{split}
\end{equation}
where $\mathsf{S}$ is the matrix
\[
\mathsf{S}=\begin{pmatrix}
C_A+C_g+C_c&-C_c\\
-C_c&C_B+C_g+C_c
\end{pmatrix}.
\]

The Hamiltonian can be rewritten for the case of two transmon qubits with frequency $\omega_A'$ and $\omega_B'$, as done in the previous sections:
\begin{equation}
\label{eqn:weakCouplTwoQqubitDir}
\begin{split}
\mathcal{H}=&\frac{\hbar\omega_A'}{2}\sigma_A^z+\frac{\hbar\omega_B'}{2}\sigma_B^z+\mathsf{S}^{-1}_{12}\lambda_A'\lambda_B'\sigma_A^y\sigma_B^y+\sum_{k}\hbar\omega_k a_k^\dagger a_k\\
&+iC_g\sum_{k}\sqrt{\frac{\Re[Z(\omega_k)]\hbar\omega_k\Delta\omega}{\pi}}\left(\lambda_A'(S^{-1}_{11}+S^{-1}_{12})\sigma_A^y+\lambda_B'(S^{-1}_{21}+S^{-1}_{22})\sigma_B^y\right)(a_k^\dagger-a_k),
\end{split}
\end{equation}
where, as before, $\lambda_A'$ and $\lambda_B'$ are constants depending on the intrinsic values of the energies of the transmon qubits \cite{Koch2007a}, given by the substitution of $p_j$ by $\sigma_j^y$, through $p_j\approx \lambda_j'\sigma_j^y$.
In this scenario, the Hamiltonian is not diagonal in the canonical basis of the qubits A and B, and the master equation must be derived by finding the eigenmodes of the system. This makes the problem and the final expression of the master equation more complex. We refer the interested reader to Ref.~\cite{Cattaneo2019} for its specific structure.

\subsection{Two qubits, separate baths}
\label{sec:separate}

\begin{figure}
\center
\includegraphics[scale=0.55]{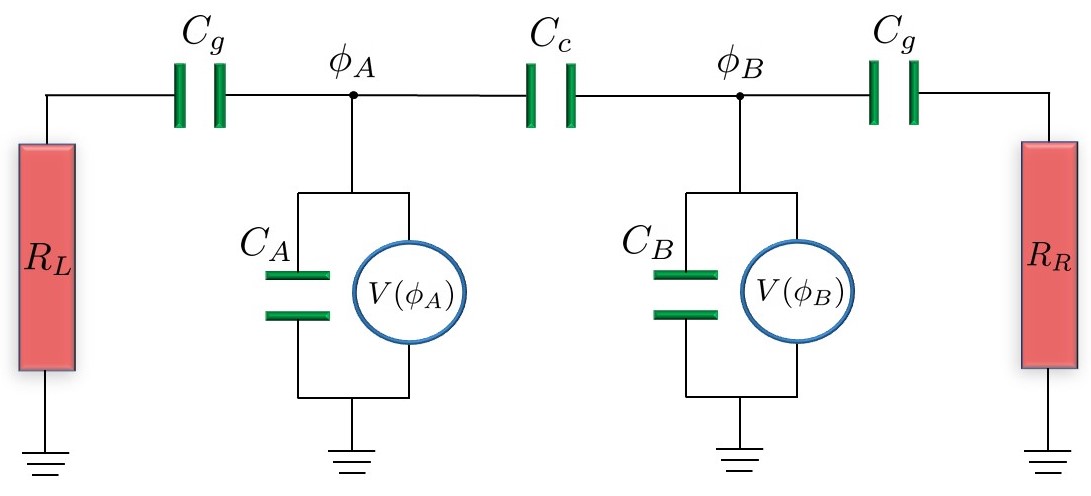}
\caption{Two qubits $A$ and $B$ coupled through a capacitor with capacitance $C_c$, and independently coupled to two resistors $R_L$ and $R_R$, playing the role of separate baths, at least if the coupling capacitances $C_g$ are small. The resistors are described according to the Foster's first form, and the node fluxes of the internal degrees of freedom (in the present picture not shown) are the ones depicted in Fig.~\ref{fig:firstFoster}.}
\label{fig:separateBaths}
\end{figure}
Finally, let us analyse the case in which we couple two different resistors to two qubits, and then couple the qubits at the center of the circuit as depicted in Fig.~\ref{fig:separateBaths}. The qubit-bath coupling capacitors have the value $C_g$, while the qubit-qubit coupling capacitor $C_c$. We find the Hamiltonian of the system in the weak coupling limit, under which all the coupling capacitors are weak compared to the capacitances of the qubits: $C_g,C_c\ll C_A,C_B$. Under this condition, the circuit in Fig.~\ref{fig:separateBaths} may be seen as the chain of capacitively coupled LC circuits discussed in Sec.~\ref{sec:capCoupling}. In analogy with that scenario (see the discussion in Appendix~\ref{sec:appendixTwoQSep}), we can thus write the Hamiltonian in the weak coupling limit as:
\begin{equation}
\label{eqn:HamtwoQseparate}
\begin{split}
\mathcal{H}=&\frac{\mathsf{S}^{-1}_{11}}{2}p_A^2+V(\phi_A)+\frac{\mathsf{S}^{-1}_{22}}{2}p_B^2+V(\phi_B)+\mathsf{S}^{-1}_{12}p_Ap_B+\sum_{\alpha=1}^\infty \hbar\omega^{(L)}_\alpha (a_\alpha^{(L)})^\dagger a_\alpha^{(L)}\\
&+\sum_{\beta=1}^\infty \hbar\omega^{(R)}_\beta (a_\beta^{(R)})^\dagger a_\beta^{(R)}+iC_g\sum_{\alpha=1}^\infty\sqrt{\frac{\Re[Z_L(\omega_\alpha)]\hbar\omega_\alpha\Delta\omega}{\pi}}\mathsf{S}^{-1}_{11}p_A((a_\alpha^{(L)})^\dagger-a_\alpha^{(L)})\\
&+iC_g\sum_{\beta=1}^\infty\sqrt{\frac{\Re[Z_R(\omega_\beta)]\hbar\omega_\beta\Delta\omega}{\pi}}\mathsf{S}^{-1}_{22}p_B((a_\beta^{(R)})^\dagger-a_\beta^{(R)}),\\
\end{split}
\end{equation}
where $\mathsf{S}$ is the matrix
\[
\mathsf{S}=\begin{pmatrix}
C_A+C_g+C_c&-C_c\\
-C_c&C_B+C_g+C_c
\end{pmatrix},
\]
and we have neglected any contribution beyond the first order of the weak coupling approximation. The bath frequencies are given by $\omega_{\alpha}(\omega_{\beta})=\Delta\omega\alpha(\Delta\omega\beta)$, and taking the limit $\Delta\omega\rightarrow 0$ we recover an Ohmic spectral density as in Eq.~\eqref{eqn:spectralDensitySingleQ}.

Considering A and B as transmon qubits with frequency $\omega_A'$ and $\omega_B'$, we have:
\begin{equation}
\label{eqn:HamtwoQseparateQubit}
\begin{split}
\mathcal{H}=&\frac{\hbar\omega_A'}{2}\sigma_A^z+\frac{\hbar\omega_B'}{2}\sigma_B^z+\mathsf{S}^{-1}_{12}\lambda_A'\lambda_B'\sigma_A^y\sigma_B^y+\sum_{k} \hbar\omega^{(L)}_k (a_k^{(L)})^\dagger a_k^{(L)}\\
&+\sum_{k'} \hbar\omega^{(R)}_{k'} (a_{k'}^{(R)})^\dagger a_{k'}^{(R)}+iC_g\sum_{k}^\infty\sqrt{\frac{\Re[Z(\omega_k)]\hbar\omega_k\Delta\omega}{\pi}}\lambda_A'S^{-1}_{11}\sigma_A^y((a_k^{(L)})^\dagger-a_k^{(L)})\\
&+iC_g\sum_{{k'}}\sqrt{\frac{\Re[Z(\omega_{k'})]\hbar\omega_{k'}\Delta\omega}{\pi}}\lambda_B'S^{-1}_{22}\sigma_B^y((a_{k'}^{(R)})^\dagger-a_{k'}^{(R)}),
\end{split}
\end{equation}
where $\lambda_A'$ and $\lambda_B'$ are constants depending on the intrinsic values of the energies of the transmon qubits \cite{Koch2007a}, given by the substitution of $p_j$ by $\sigma_j^y$, through $p_j\approx \lambda_j'\sigma_j^y$ (see the discussion for the previous examples). In the weak coupling limit $C_c,C_g\ll C_A,C_B$, we can set $S_{12}^{-1}\approx C_c/(C_A C_B)$, $S_{11}^{-1}\approx C_A^{-1}$, $S_{22}^{-1}\approx C_B^{-1}$.

Eq.~\eqref{eqn:HamtwoQseparateQubit} is the Hamiltonian of two coupled qubits interacting with separate baths. This is a fundamental model, for instance, for the study of heat transport in quantum thermodynamics \cite{Anders2016}. The requirement $C_c\ll C_{A,B}$ induces a weak coupling between the qubits (in the circuit Hamiltonian, $\mathsf{S}_{12}^{-1}\ll\mathsf{S}_{11}^{-1},\mathsf{S}_{22}^{-1}$). This allows us to solve the system dynamics by employing a local master equation \cite{Cattaneo2019,Hofer2017}, in which the dissipation is computed locally on each qubit, and the interaction between them appears only in the unitary part of the master equation. If $\rho_{AB}(t)$ is the two-qubit state at time $t$, its dynamics is given by:
\begin{equation}
\label{eqn:masterEqSep}
\begin{split}
\frac{d}{dt}\rho_{AB}(t)=&-\frac{i}{\hbar}[H_S+H_{LS},\rho_{AB}(t)]+\underbrace{\sum_{j,k=A,B}\Gamma_{j}^\downarrow (\beta )\left(\sigma_j^-\rho_{AB}(t)\sigma_j^+-\frac
{1}{2}\{\sigma_j^+\sigma_j^-,\rho_{AB}(t)\}\right)}_{\textnormal{emission}}\\
&+\underbrace{\sum_{j,k=A,B}{\Gamma}^\uparrow_{j}(\beta )
\left(\sigma_j^+\rho_{AB}(t)\sigma_j^- 
-\frac{1}{2}\{\sigma_j^-\sigma_j^+,\rho_{AB}(t)\}\right),}_{\textnormal{absorption}}
\end{split}
\end{equation}
where $H_S=\frac{\hbar\omega_A'}{2}\sigma_A^z+\frac{\hbar\omega_B'}{2}\sigma_B^z+C_c/(C_A C_B)\lambda_A'\lambda_B'\sigma_A^y\sigma_B^y$ is the system Hamiltonian containing the unitary interaction between the qubits, while $H_{LS}$ is a frequency-shifting Lamb-shift term \cite{BreuerPetruccione,Cattaneo2019}. The local coefficients $\Gamma_{j}^{\downarrow\uparrow}(\beta )$ depend on the temperature of the bath coupled to the qubit $j$, and on the corresponding Ohmic spectral density $J_j(\omega)$ obtained according to Eq.~\eqref{eqn:spectralDensitySingleQ}; analogously to Eq.~\eqref{eqn:decayRates}, they are given by:
\begin{equation}
\label{eqn:decayRatesLocal}
\begin{split}
&\Gamma_{j}^{\downarrow}(\beta )=\left(\frac{\lambda_j' C_g}{\hbar C_j}\right)^2 S_{V_jV_j}(\omega_j')=\pi J_j(\omega_j)\left(\coth\frac{\beta\hbar\omega_j}{2}+1\right),\\
&\Gamma_{j}^{\uparrow}(\beta )=\left(\frac{\lambda_j' C_g}{\hbar C_j}\right)^2 S_{V_jV_j}(-\omega_j')=\pi J_j(\omega_j)\left(\coth\frac{\beta\hbar\omega_j}{2}-1\right).\\
\end{split}
\end{equation}

\section{Experimental methods for control of dissipation}
\label{sec:control}

For real-life applications, the simplest dissipative circuit elements can be realized either by using resistive components or lossy transmission lines. In both cases the experimentalist has the choice of integrating them on-chip or to realize separately the dissipative unit from components. In the first case, metal evaporation and e-beam lithography in the same process as the qubits are the methods of choice. Also several types of resistors are commercially available, and in general they can be utilized at low temperatures. Commercial types of resistors include wirewound resistors, metal-foil resistors, thin-film (tantalum, nickel-chromium, carbon-boron), carbon-based resistors, and ceramic-metal powder resistors. 
A semi-infinite transmission line is another way of realizing a resistive element. This type of model can be written down based on the standard procedure of quantization of transmission lines, and the spectral power density can be calculated. Alternatively, the spectral power density can be obtained directly from our formalism by the following argument: a semi-infinite transmission line can be constructed - both in theory and in practice - by terminating the line with a matched impedance. If the line has impedance $Z_{0}=\sqrt{L'/ C'}$, where $L'$ and $C'$ are the inductance and capacitance per unit length, then we add a load of the same value $Z_{0}$ (typically 50 $\Omega$). From the elementary classical treatment of transmission lines, we know that in this case the impedance seen at the input of the line is also $Z_{0}$. By applying Eq. (\ref{eqn:spectrumResistor}) we find $S_{VV} = 2\hbar \omega Z_{0}\mathcal{N}(\omega )$, which is the same as the result obtained from the quantization procedure.
It is also possible to create filtered spectral spectral power densities using transmission lines. For example, consider a transmission line of length $l=\lambda/2$ with a shortcut at one end. An analysis of the impedance as seen from the other end shows that the device can be approximated around the resonance as an RLC series circuit with $L=lL'/2$, $C=lC'/2$, and $R=lR'/2$, where $L'$, $C'$ and $R'$ are respectively the inductance, capacitance, and resistance per unit length. Here the factor $1/2$ is due to the formation of standing waves in the line, with voltages and currents varying from zero to the maximum value.

However, fabrication of a standard resistor or the transmission line are not  the only methods to produce, and especially to  control dissipation. Indeed, to date various forms of dissipation engineering have been proposed or realized in experiments, involving a variety of techniques, from passive methods to active techniques employing tunable devices and microwave-assisted dissipation.  Irrespective of the physical realization (typical examples included below are low-Q coplanar waveguide resonators, Josephson junctions, and metallic resistors fabricated on-chip), it is important to understand that the dynamics of the electromagnetic collective degrees of freedom is, generically, the same.
For applications such as quantum computing, at first sight the use of any dissipative circuit element is something to avoid, as one aims at high coherences. However, if controlled, dissipation can serve as a way for high-speed preparation of states that are useful for quantum computing, for example for initializing the qubits in the ground state $|00000 .. \rangle$ at the beginning of a quantum algorithm, or for dumping the excess photons in the resonator used for readout. The latter situation is useful to eliminate the dephasing of the qubit caused by the parasitic photons from the resonator and to reduce the measurement time and the lag between successive measurements. In quantum computing, high-fidelity and fast measurements are essential in running error-correction codes - and more generally for feedback experiments (where the results on measurements performed on ancillas are used to correct the state of the logical qubits).

\subsection{Passive control and circuit design}

In mesoscopic physics, dissipative elements have traditionally played a role mostly in realizing filters and ensuring good thermalization. But recently they took a more prominent role \cite{PekolaQT} as thermal reservoirs in quantum thermodynamics -- the study of fluctuations and small-scale thermal engines at the single-quantum level. For instance, they have been key ingredients for some proposals and realizations of calorimeters in circuit QED \cite{Gasparinetti2015,Karimi2018,Karimi2020}, with applications to the measurement of work in a quantum system \cite{Pekola2013a,Viisanen2015a}. Moreover, they play a major role in the study of quantum heat transport \cite{Campisi2015,Campisi2017}, in the proposal for a quantum heat switch \cite{Karimi2017} and an Otto refrigerator \cite{Karimi2016}, in an experimental demonstration of quantum-limited heat conduction \cite{Partanen2016}, and in the creation of a heat sink for quantum circuits \cite{Partanen2018}. In the experiments with a qubit operated as a heat valve \cite{Ronzani2018} as well as in the demonstration of heat flux rectification \cite{Senior2019} and of electrostatic control of 
 radiative heat transfer \cite{Maillet2020}, the reservoirs were fabricated as Cu-metal resistors on the same chip.  The resistors can be Joule-heated, thus providing the temperature difference needed to study heat transport phenomena, and, in the future, to build mesoscopic-scale thermal engines. The slow electron-phonon relaxation is an advantage for the use of resistors as thermal baths, since they do not require excessive amounts of energy that would produce heating of the substrate and of the other on-chip components. In these experiments, the SIN structures have a resistance of 22 $k\Omega$ and they can be heated by applying a DC voltage larger than the superconducting gap.

\paragraph{Filtering}

\begin{figure}
\centering
\subfloat[]{%
  \includegraphics[scale=0.28]{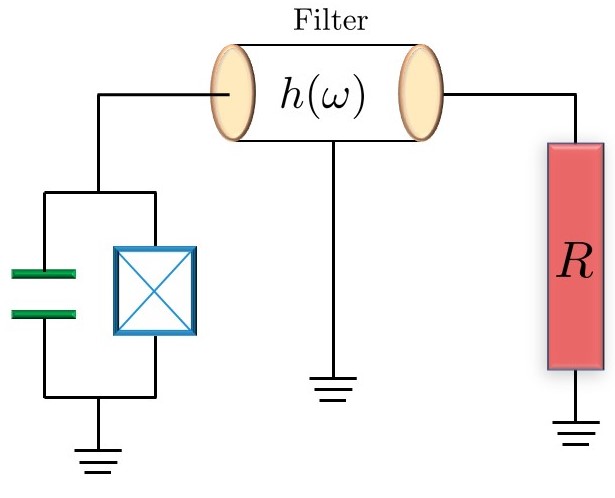}%
}\qquad
\subfloat[]{%
  \includegraphics[scale=0.23]{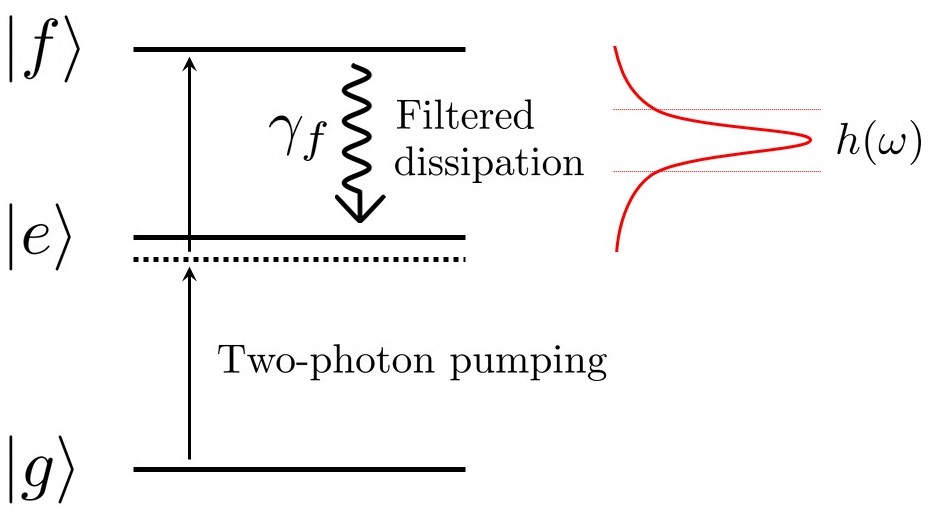}%
}\qquad
\subfloat[]{%
  \includegraphics[scale=0.23]{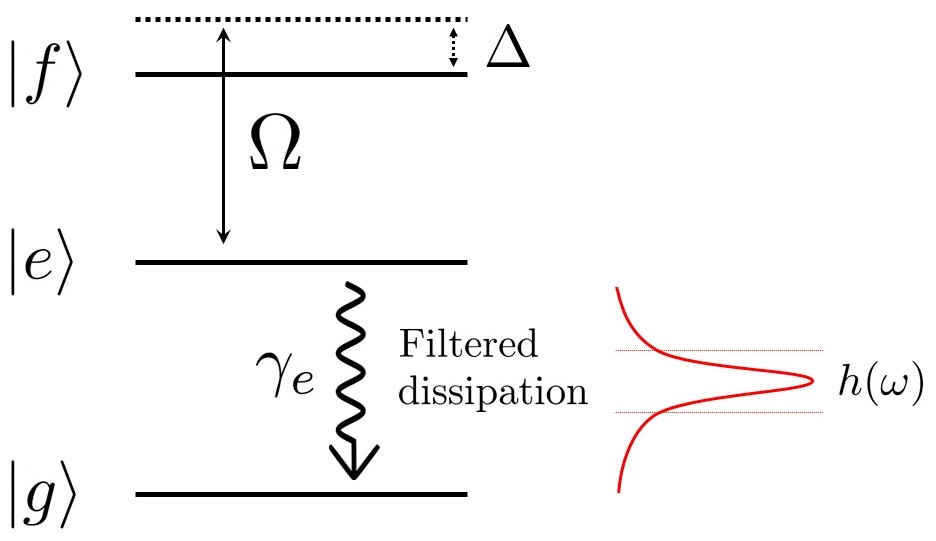}%
}
\caption{Generic schemes for dissipative control of quantum circuits. (a) Filtering of the thermal noise of a resistor coupled to a qubit.
(b) Stabilization of single-qubit excited state $|e\rangle$ (population inversion) by nearly-resonant filtered coupling to a dissipating reservoir in conjunction with a continuous tone resonant applied to the $g$ -- $f$ transition by a 2-photon process. (c) Creation of an effective non-Hermitian Hamiltonian in the subspace $\{|e\rangle , |f\rangle \}$ by enhancing the decay rate of level $e$ in conjunction with a microwave drive applied to the
$e$ -- $f$ transition.}
\label{fig:control}
\end{figure}

In general, if $X(t)$ is a random variable at the input of a linear time-invariant circuit with impulse response $h(t)$ and output $Y(t)$, then the power spectral density of the variable $Y(t)$ is $R_{YY}(\omega ) = |h(\omega )|^{2}S_{XX}(\omega )$. Here $h(\omega )$ is the transfer function - the Fourier transform of the impulse function $h(t)$, see Fig. \ref{fig:control}. One such very useful linear time-invariant device is a band-pass filter, which allows us to shape the spectrum of the noise in very useful ways. We have seen already an example of this in Section 4.1.1, where we analysed in detail the effect of an LC filter. 
An ideal bandpass filter with low-pass frequency $\omega_{\rm L}>0$ and high-pass frequency $\omega_{\rm H}>0$ has the transfer function
\begin{equation}
h(\omega ) = \begin{cases} 1 & {\rm if}~~  \omega_{\rm L} < |\omega | < \omega_{\rm H}, \\ 0 & {\rm otherwise}.
\end{cases}
\end{equation}
Since we are interested in voltage power spectral density,
the shaping action of a filter with transfer function $h(\omega )$ is given by
\begin{equation}
S_{VV}(\omega ) \rightarrow |h(\omega )|^{2}S_{VV}(\omega ).
\end{equation}
 We assume here that the quantum system at the output of the filter, which acts as a load,  has a very large impedance in the bandwidth region, therefore its effect is neglected. It is important to notice that the action of the filter applies both to negative and positive frequencies $\omega$; that is, the filter works in the same way both when the system absorbs and emit energy. The bandwidth $\omega_{\rm H}-\omega_{\rm L}$ of the filter can be designed narrow enough such that it ensures selective coupling with the quantum system of interest (with only one of the transitions for a transmon qubit or with the resonant frequency if it is a superconducting resonator). 
 
 Two immediate applications of this scheme are the realization of Purcell filters and the filtering of thermal reservoirs in quantum thermodyamics. The readout resonator used in circuit QED to measure the qubits has to satisfy contradictory requirements: it needs to have a high quality factor so that the qubit is not damped and at the same time it needs to allow photons to pass in and out such that the measurement is performed in an as short measurement time as possible. To protect the qubit, it is easy to design circuits such that the qubit frequency is placed on the tail of the response curve of the dissipative element, as far as possible from the maximum. However, additional filtering offers an even better protection, and also enables simultaneous measurement of several qubits by multiplexing.  For example, single-pole bandpass filtering of the signal from the resonator is a good compromise for the requirements above, preserving a  high $T_1 > 10$ $\mu$s and allowing a high-fidelity (99.8\%) and at the same time fast (120 ns) measurement of the qubit \cite{Martinis2014}.  In this case, the filter (with quality factor of about 30) was realized on-chip as a single $\lambda /4$ coplanar waveguide resonator. The readout resonators are placed in the passband, thus 
preventing the dissipative modes to dampen the qubit and at the same time allowing for a significant outcoupling for the modes around the resonator frequencies. Another design uses two  $\lambda /4$ resonators \cite{Reed2010} to achieve a zero real admittance environment at the qubit frequency, thus preventing the spontaneous decay through the outcoupled resonator modes. In this case, an improvement in $T_1$  by a factor of 50 was observed at a qubit frequency of 6.7 GHz --  at the same frequency at which the transmission through the resonator followed by the Purcell filter showed a 30 dB drop.  Finally, in quantum thermodynamics the use of resonators placed between resistors and the qubit and acting as spectral filters is a necessary requirement in order to achieve selective thermalization of the qubit with either one of the reservoirs \cite{Ronzani2018,Senior2019}.

 \paragraph{Tuning of dissipation}
 
Tunability is an important property of a dissipative device, opening up the way for many applications. The easiest way to achieve this is to incorporate a SQUID into the device, and change the critical current by the use of an external magnetic field. 
Proposals along this line include the dissipator as a frequency-tunable Josephson junction with small quality factor ($\approx 100$) \cite{Wong_2019}. A design where two resonators, one with high quality factor and another one tunable and coupled to the ground by a resistor, has been studied in Refs. \cite{Jones2013,Partanen2018}. It was observed that the loaded quality factor of the high-Q cavity ($10^5$) can be tuned down by a factor of almost 100. In this experiment the resonators were made of Nb, while the resistor was made of Cu (R=375 $\Omega$). Another scheme for creating wideband tunable dissipative elements 
has been studied \cite{Rastelli2018}; the device consists of a double chain of SQUIDs capacitively coupled to each other, which sustains modes with wide dispersion. Dissipation in this circuit is realized by the capacitive coupling to a transmission line of both of the SQUID chains. These devices can be employed to efficiently initialize the state of a superconducting qubit \cite{Tuorila2017}, a well-known fundamental task in circuit QED.  Further experiments could use photon-assisted tunneling in an NIS junction to cool a device such as a qubit or a resonator \cite{Tan2017}. \mar{Present practical methods used in more advanced quantum processors \cite{McEwen2021} for initialization also make use of tuning of dissipation, based on the observation that it is more convenient experimentally to tune the qubit rather than the bath. In a standard transmon qubit coupled to the readout coplanar waveguide cavity, the qubit (with a higher frequency than  the cavity) is adiabatically tuned down towards resonance with the cavity, which swaps the excess population from the excited states of the qubit into the cavity. The qubit is then held at a frequency lower than the cavity. Meanwhile, the excitations transferred to the cavity are fast dissipated into the environment. Note that the cavity, with decay rate $\kappa \approx 1/(45~{\rm ns})$, relaxes much faster than the qubit ($T_{1}\approx 14~\mu$s), therefore it truly plays the role of a dissipative environment. After the cavity is sufficiently depleted (typically 200 ns), the qubit is returned diabatically to the operating frequency. During this last step, it crosses again the cavity, but it will not get populated since the cavity is empty. The whole process lasts only 280 ns and achieves a ground state with error below 0.5\%.}

\mar{Finally, certain tasks in quantum thermodynamics, for example related to the operation of proposed Stirling engines \cite{Sina2020}, require a fast-tunable coupling between a qubit and a reservoir. For this, various designs for tunable coupling have been developed in the context of making two-qubit gates in quantum computing \cite{Chen2014,Mundada2019}, which can be easily adapted to coupling a qubit to a bath.}

\subsection{Active control of dissipation by coherent drives}

An interesting recent development is the realization of schemes where dissipation is controlled by coherent microwave drives combined with filtering. 

For example, a reset method using a Josephson photomultiplier as a photon counter was demonstrated for a transmon qubit \cite{Opremcak1239}.
The Josephson photomultiplier consists of a capture cavity with relatively low quality factor ($\approx 1300$) capacitively coupled to a Josephson junction placed in a rf-SQUID and biased near the phase-slip critical flux. In this experiment, the dissipating device was fabricated as a standalone device from Al on high-resistivity Si, packaged in an Al box, and connected by a coaxial wire to the measuring resonator of the qubit. 
With this device, a deterministic reset of the measuring cavity can be realized by a two-pulse sequence. The first is a fast flux pulse that resets the photomultiplier itself by putting it in a single-well potential configuration. The second pulse puts the qubit cavity in resonance with the capture cavity (and nearly-resonant with the qubit cavity), with the result of depleting both cavities of spurious photons. Ramsey interference fringes show a clear increase in visibility as the depletion time is increased. \mar{Another method based on two microwave drives was demonstrated in Ref. \cite{Geerlings_2013}.}

In the same vein, a scheme for initialization or reset can be constructed by using a parametric modulation of the qubit frequency $\omega (t) = \omega + A \cos (\nu t)$, which creates sidebands \cite{Tencent2021}. When the n's sideband becomes resonant with the readout resonator $\omega_{d}$ (typically lower than $\omega$),
$n \nu + \omega_{d}=\omega$, transfer of population can occur. The microwave-induced coupling  \cite{paraoanu2006} between the qubit and the resonator results in the interaction Hamiltonian between the qubit and the lossy cavity (written in the rotating frame)
\begin{equation}
    H_{ge-d}=g J_{n}\left( \frac{A}{\nu}\right) \left( \sigma_{ge} d^{\dag} + 
    \sigma_{ge}^{+}d\right),
\end{equation}
where $\sigma_{ge} = |g\rangle \langle e|$, $d$ is the mode of the lossy cavity (in this case the readout cavity), $g$ is the bare (unmodulated) qubit-resonator interaction, and $J_{n}$ is the Bessel function of first kind. Swapping the population between the qubit and the resonator, due to the results in the suppression of the residual excited population to 0.08 \% in only 34 ns \cite{Tencent2021}. This type of interaction has multiple application in circuit QED, such as entanglement production \cite{Li2008},
realization of cross-resonant qubit-qubit gates
\cite{chow2011,chow2012}, 
 motional averaging effects \cite{Li2013}, and Landau-Zener-St\"uckelberg-Majorana interference \cite{Silveri_2015,MikaS2019}.

Selective coupling to dissipative elements is not limited to two-level systems. Transmon circuits have higher energy levels that can be accessed experimentally. In Fig. \ref{fig:control} (b) we show such a configuration, where a transmon is truncated to the first three-levels $|g\rangle$, $|e\rangle$, and $|f\rangle$. The $e-f$ transition is coupled to a dissipative element via a filter ensuring that the $g-e$ transition is off-resonant, therefore not affected. \mar{If the $g-f$ transition is pumped resonantly at a frequency of half the sum of the $g-e$ and $e-f$ transition frequencies}, two-photon processes promote the population to state $f$, from which it decays fast to state $e$. This results in population inversion in the $\{|g\rangle , |e\rangle \}$ manifold, stabilizing the excited state $|e\rangle$.
That this is the case can be seen by considering the truncation of the Hamiltonian in the  $\{|e\rangle , |f\rangle \}$ manifold, which in the rotating-wave approximation is
\begin{equation}
    H_{ef-d} = g \hbar (\sigma_{ef} d^{\dag} + \sigma_{ef}^{+} d) ,
\end{equation}
where $\sigma_{ef} = |e\rangle \langle f|$, $\sigma_{ef}^{+} = |f\rangle \langle e|$ and $g$ is the coupling with the dump cavity $d$. Since $d$ is a highly dissipative mode, the stabilized state $|ss\rangle$ will necessarily have the cavity in the vacuum state $|ss\rangle |0\rangle_{d}$. Enforcing that this is an eigenvalue of the above Hamiltonian results in the condition $\sigma_{ef}|ss\rangle = 0$ or $|ss\rangle = |e\rangle$.
Note also that the two-photon process is sufficiently off-resonant with both the $g-e$ and $e-f$ transitions. \mar{Another scheme that uses three levels and is based on a microwave drive coupling the states $|f,0\rangle$ and $|g,1\rangle$ of a transmon
coupled to a dissipative broadband resonator with $\kappa = 56.52$ MHz has achieved a fast reset (500 ns) resulting in 0.2\% residual population \cite{Magnard2018}. Subsequently, it was shown that the read-out resonator can serve as the dissipative dump resonator, therefore reducing considerably the number of circuit elements needed \cite{Egger2018}.} This concept can also be generalized to multiple qubits. For example, similar ideas \cite{Mirrahimi2013} were used to \mar{create arbitrary superpositions of the ground and excited state \cite{Murch2012},} to 
autonomously stabilize two transmon qubits in a three-dimensional cavity in a Bell state \cite{Shankar2013}, and to produce a stable bosonic Mott insulator state in an array of qubits \cite{Schuster2019}.

Dissipation, when balanced with amplification, gives rise to a generalization of standard unitary evolution known as $\mathcal{PT}$-symmetric quantum mechanics \cite{Bender1998}. Experiments on exceptional points and broken $\mathcal{PT}$-symmetry have been realized on many experimental platforms. In circuit QED, a digital simulation of single-qubit and two-qubit non-Hermitian dynamics has been realized on the IBM superconducting processor \cite{Dogra2021}. It is possible to emulate the physics of these Hamiltonians in a dedicated transmon in a three-dimensional cavity setup \cite{Murch2019}, following the same principles of noise filtering as presented above, but with the interesting difference that now the dissipative channel is coupled selectively to the $g-e$ transition, see Fig. \ref{fig:control} (c). Experimentally, this is done by increasing the decay rate \mar{$\Gamma_{e}$} through the insertion of a mismatching element between the cavity and the parametric amplifier. This produces standing waves in the transmission line, resulting in different density of modes at different frequencies and therefore different Purcell rates. Since the transmon is tunable, it is possible to match the $g-e$ transition frequency to a frequency region with large density of states 
resulting in a dissipation rate \mar{$\Gamma_{e} = 5.25$} $ \mu s^{-1}$, more than one order of magnitude larger than \mar{$\Gamma_{f} = 0.25$} $\mu s^{-1}$.
The condition \mar{$\Gamma_{e}\gg \Gamma_{f}$} has as consequence an effective dynamics on the $\{|e\rangle , |f\rangle \}$ manifold governed by
\begin{equation}
    H_{\rm ef} = \hbar (\Delta - i \gamma_{e}/2)|e\rangle \langle e| + 
    \frac{\hbar \Omega}{2} (|e\rangle \langle f| + |f\rangle \langle e|) ,
\end{equation}
where $\Delta$ is the detuning of the drive from the $e-f$ transition and $\Omega$ is the coupling strength.
Note that in this case we are not interested in the steady-state imposed by dissipation (which, because of the conditions above, is trivially just the ground state for the transmon), but on the dynamics on a timescale shorter than \mar{$1/\Gamma_{e}$.}

One recognizes here the paradigmatic non-Hermitian Hamiltonian for qubits, with the dissipative term on the diagonal and a $\sigma_{x}$-coupling in the off-diagonal elements. Both in the case of  digital simulation \cite{Dogra2021} and in the case of emulation by a three-level system, postselection plays a key role: each experimental run ends with a measurement and only the data corresponding to the correct manifold is retained. In  the digital simulation \cite{Dogra2021} this is flagged by the ancilla qubit being projected in the state 1, while in the three-level system emulation this is enabled by the single-shot quantum tomography, where the occurrences of projection to the ground state are eliminated.

With continuous variables, such an active method can be employed for the stabilization of highly squeezed states (exceeding the standard 3 dB achieved by parametric drive with a single pump) in a cavity, by using an additional dump cavity \cite{huard}. The concept is similar to the case presented above in Fig. \ref{fig:control}, with the difference that now we have a harmonic oscillator instead of a transmon. In addition, in this experiment the two cavities were coupled to each other by a Josephson ring modulator, which enables the modulation of the couplings by applied microwave tones. The cavity had resonant frequency $\omega_{c}$ and decay rate $\kappa_{c}$, while the dump cavity had $\omega_{d} >\omega_{c} $ and the decay rate $\kappa_{d} \gg \kappa_{c}$ ($\kappa_{d}/\kappa_{c} = 200$ in the experiment). By pumping the Josephson ring modulator simultaneously at the difference frequency $\omega_{-} = \omega_{d} - \omega_{c}$ and at the sum frequency $\omega_{+} = \omega_{d} + \omega_{c}$
one obtains by standard methods involving the rotating-wave approximation (RWA) $H=H_{-} + H_{+}$, where
$H_{-}= \hbar g_{-}(cd^{\dag} + c^{\dag}d)$ is an effective beam-splitter Hamiltonian and $H_{+}= \hbar g_{+}(c^{\dag} d^{\dag}  + c d)$ is an effective parametric down-conversion Hamiltonian. The truly interesting feature emerges when the high damping of mode $\rm d$ is considered. Indeed, arranging the phases and amplitudes of the pumps such that $g_{\pm}$ are positive and real with $g_{-} >g_{+}$, the full effective Hamiltonian can be rewritten as a beam-splitter Hamiltonian between the dump mode and a Bogoliubov mode 
\begin{equation}
H_{c-d} = \hbar G(\mathcal{B}^{\dag} d + \mathcal{B} d^{\dag}),\label{eq:inspec}
\end{equation}
where $G = \sqrt{g_{-}^2 - g_{+}^{2}}$ and $\mathcal{B}$ is the Bogoliubov mode of the cavity $c$, defined by $\mathcal{B} = c \cosh r  + c^{\dag} \sinh r $ with squeezing coefficient $r= \tanh^{-1}(g_{+}/g_{-})$. 
Now, if $\kappa_{\rm d} \gg \kappa_{\rm c}$, dissipation forces the dump resonator into the vacuum state $|0\rangle_{\rm d}$. As a result, by inspecting the Hamiltonian Eq. (\ref{eq:inspec}), we conclude that the steady-state $|ss\rangle_{\rm c}$ of the cavity $\rm c$ must be the vacuum state of the Bogoliubov mode $\beta|ss\rangle_{\rm c}=0$, which is a squeezed state with squeezing parameter determined by the ratio $g_{+}/g_{-}$. In principle, one can increase $r$ to very large values by having $g_{+}$ nearly equal to $g_{-}$. This however reduces the coupling $G$, so in practice an optimal value has to be found. In the experiment \cite{huard}, an optimal squeezing slightly above 8 dB was obtained. Another way of seeing this is by analysing the time it takes to reach the steady state, which  is of the order of $\kappa_{d}/G^2$. This time increases as we aim at higher squeezing factors by increasing $g_{+}$ to approach from below $g_{-}$, because $G$ decreases as well in this situation.

\mar{
Along the same lines, an important application of these concepts is the realization of a single-photon detector \cite{Flurin_2020}. This device uses two cavities (a buffer cavity where the photon to be detected is first stored, and a second highly-dissipative waste cavity), with a coupling transmon qubit in-between. Denoting the buffer and waste annihilation operators by $b$ and $w$, and the qubit lowering operator by $\sigma$, it is possible to derive an effective Hamiltonian
\begin{equation} \label{eq:singleph}
H_{eff}=\hbar g_{3}(b\sigma^{\dag}w^{\dag}  + b^{\dag}\sigma w ),\end{equation}
where $g_3$ is an effective three-wave mixing rate, activated by the parametric pumping of the transmon. This Hamiltonian converts a photon incident on the buffer into a qubit excitation and a waste photon.
To realize the effective Hamiltonian Eq. (\ref{eq:singleph}), the transmon is pumped at a frequency $\omega_p$ such that $\omega_{p} = \tilde{\omega}_{ge} + \omega_{w}^{e}-\omega_{b}^{g}$, where $\tilde{\omega}_{ge}$ is the ac-Stark shifted qubit frequency, $\omega_{w}^e$  is the waste-cavity frequency with the qubit being in the excited state, and $\omega_{b}^g$ is the buffer cavity frequency with the qubit in the ground state. When the buffer contains one photon, the pump provides the missing energy of transforming that photon into an excitation on the qubit and a photon in the waste. Now, if the dissipation rate $k_w$ of the waste is such that $|g_{3}|\ll k_{w}$, then the waste photon is eliminated fast into the environment, while the qubit remains in the excited state. By adiabatically eliminating the waste, a non-local and non-linear loss operator $2|g_{3}|k_{w}^{-1/2}b\sigma^{\dag}$ is obtained, which shows that the qubit goes into the excited state conditioned on the presence of a photon in the buffer. Finally, the excited state of the qubit is detected by a standard single-shot dispersive technique. This detector has
a detection efficiency of 58\% and a  dark count rate of 1.4 photons per millisecond. }


\section{Conclusions and perspectives}
\label{sec:conclusions}

Modeling of electrical circuits comprising resistors and the other dissipative elements has experienced a new momentum recently, driven by burgeoning applications in circuit quantum electrodynamics. Here we have reviewed some of the standard models in the field, and we outlined the main techniques for coupling resistors with qubits and controlling dissipation. This opens a number of interesting perspectives  for future works, as many theoretical proposals rely on engineering thermal baths acting on quantum devices.

In quantum information and quantum computing, as we have already seen, engineered reservoirs provide an advantage with respect to
adiabatic state preparation, which is slow - since it should be slower compared to the energy gap. Also, in dissipation engineering of quantum states the final state is protected from perturbations – implementing in fact a form of self-correction. Tunable Markovian dissipation provides a scheme for universal quantum computation \cite{Verstraete2009,Paulisch2016} and for the characterization of information-preserving steady states \cite{Blume-Kohout2010,Albert2016}. Moreover, non-Markovian effects may provide new insights into the exchange of information between a system and the environment that surrounds it \cite{Rivas2014,Breuer2016,DeVega2017a}. Looking forward to applications in circuit QED, \mar{ideas exploring quantum optimal control and dissipative networks could allow for novel reservoir-engineering reset schemes of superconducting qubits \cite{Zhang2015, Basilewitsch2019,Zhang2020}}. Furthermore, resistors acting as thermal bath can be used to study how non-Markovianity effects affect the work done by a driving force on a superconducting qubit \cite{HamedaniRaja2018}, and to address different collective effects in a pair of transmon qubits immersed in a common bath \cite{Cattaneo2020}. \mar{Besides, they have been used to theoretically explore quantum phase transition with dissipative frustration \cite{Maile2018}. More information can be found also in Ref.~\cite{Maile2021}.} Finally, a very recent proposal of the experimental observation of the quantum trajectory of a single microwave photon detection is based on a Cu-metal resistor that plays again the role of a thermal bath \cite{Karimi2020a}.

In superconducting circuit simulators, access to each site will enable further experiments on many-body localization, impurity (defect) dynamics  and non-equilibrium thermodynamics such as pre-thermalization, allowing for further manipulation (driving the system adiabatically from incompressible to compressive phases) of the many-body photonic state. There are theoretical proposals for realizing a photonic Bose-Einstein condensate in strongly driven 
microcavity arrays using two interacting superconducting qubits each coupled to a neighbouring cavity, where the dissipation here would be realized by the spontaneous decay of the antisymmetric mode \cite{Rabl2012}, which produces scattering of photons in low-momentum states thus stimulating condensation in the zero-momentum state. 
Exotic state preparation using engineered dissipation is possible in circuit QED \cite{PhysRevLett.123.063602}, drawing on ideas of observing photon blockade and the Mott-insulator-to-superfluid transition with polaritons \cite{Hartmann2006,Angelakis2007}. Other possibilities include dissipative preparation of topological superconductors \cite{Fazio2016}, parametric thermalization \cite{Taylor2015} and creation of compressible phases and exotic nonequilibrium processes associated with generation of entropy \cite{Iacopo2017}.

Finally, in quantum thermodynamics, proposals to realize heat engines in circuit QED will require filtering of reservoir noise together with precise control of the operation point of the qubit \cite{Sina2020}. Dissipation-by-measurement is also an emerging, fundamentally new approach \cite{PhysRevLett.118.260603}.

To conclude, in the near future, tailored dissipation will be more and more employed as a resource for the preparation of otherwise inaccessible or unstable many-body states and  for targeting and purifying states with high degree of correlations. Engineering a thermal bath acting on quantum devices is of great importance for both fundamental physics and technological applications. We also notice that some of the techniques we have presented, related to control of dissipation by the application of external fields, are not limited to the superconducting circuits. These concepts can be implemented in platforms such as trapped ions or ultracold gases, bringing in new physics and additional opportunities for experimental control.

\medskip
\textbf{Acknowledgements} \par 
The authors acknowledge interesting discussions with Jukka Pekola, Sabrina Maniscalco and Roberta Zambrini. M. C. would like to thank Adrián Parra-Rodriguez and Íñigo Luis Egusquiza for valuable clarifications and extensive discussions on the derivation of the circuit Hamiltonian. He would also like to thank Lorenzo Gavassino for useful suggestions, as well as Antti Vaaranta for a careful reading of the manuscript and for spotting some typos. M. C.  acknowledges funding from the Finnish Center of Excellence in Quantum Technology QTF (projects 312296, 336810), from the Mar\'{i}a de Maeztu Program for Units of Excellence in R\&D (MDM- 2017-0711), and from Fondazione Angelo della Riccia. G. S. P. would like to acknowledge 
support from the RADDESS programme (project  no. 328193) of the Academy of Finland and from the Grant No. FQXi-IAF19-06 (“Exploring the fundamental limits set by thermodynamics in the quantum regime”) of the Foundational Questions Institute Fund (FQXi), a donor advised fund of the Silicon Valley Community Foundation.

\medskip

\appendix

\section{Quantizing electromagnetic circuits}
\label{sec:quantization}
We  discuss here the original method to quantize the degrees of freedom of electromagnetic circuits, relying on the quantization of the branch fluxes of the circuital network \cite{Yurke1984,Vool2017} (for a different approach, see \cite{Ulrich2016}).

\begin{wrapfigure}{r}{0.35\textwidth}
  \begin{center}
    \includegraphics[scale=0.4]{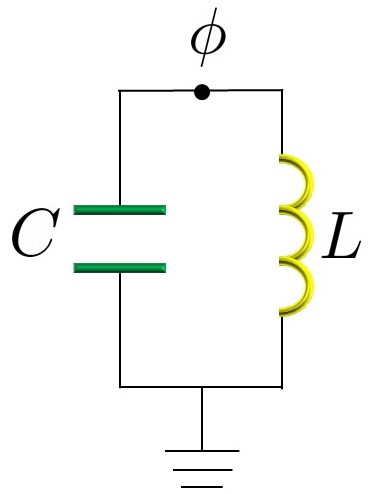}
  \end{center}
  \caption{Parallel LC circuit with corresponding node flux $\phi$.}
  \label{fig:LCcircuit}
\end{wrapfigure} 

The flux of a certain branch of the circuit is defined as the integral of the voltage difference across it:
\begin{equation}
\label{eqn:nodeFlux}
\phi(t)=\int_{-\infty}^t ds\,V(s).
\end{equation}
Straightforwardly, we have $\dot{\phi}(t)=V(t)$. Once we have defined the fluxes of all the circuit branches, we can identify the degrees of the freedom of the system by using the standard method of nodes \cite{Vool2017}. This is why the fluxes of the circuit are usually called ``node fluxes'':
the reduction given by the node method allows us to interpret the remaining relevant fluxes as associated with each node of the network, once we have defined a proper ground point. This can be observed in Fig.~\ref{fig:LCcircuit}, representing a parallel LC circuit: we only have one degree of freedom given by the node flux $\phi$, equal to the fluxes of both the capacitive and inductive branches under a suitable choice of sign. More complex networks require a more involved analysis (see \cite{Vool2017} for a deeper discussion).

Let us now focus on the LC circuit shown in Fig.~\ref{fig:LCcircuit}. Assuming the lumped element model is valid, all the inductive elements of the circuit are described by the inductor in the left branch of circuit. Therefore, making use of Faraday's law, we observe that $\phi$ is, up to a choice of sign, the magnetic flux through the inductor, related to the current $I$ running in the circuit by $\phi=L I$, where $L$ is the inductance.
We recall that the electromagnetic energies contained in a perfect capacitor and inductor respectively read $U_C=\frac{1}{2}C V^2$ and $U_L=\frac{1}{2L} \phi^2$, where $C$ is the capacitance, $L$ the inductance, $V$ the voltage across the capacitor and $\phi$ the flux through the inductor. Then, we can consider $U_C$ as the kinetic energy and $U_L$ as the potential energy of the system, and we can introduce the Lagrangian of the parallel LC circuit as:
\begin{equation}
\label{eqn:lagrangianLC}
\mathcal{L}=\frac{1}{2}C\dot{\phi}^2-\frac{1}{2L}\phi^2.
\end{equation}

One can verify the validity of this Lagrangian by writing the Euler-Lagrange equations and observing that they correspond to the proper Kirchoff's circuit laws. We can now obtain the Hamiltonian by finding the momentum and applying the Legendre transformation:
\begin{equation}
\label{eqn:hamiltonianLC}
\begin{split}
&Q=\frac{\partial\mathcal{L}}{\partial\dot{\phi}}=C\dot{\phi},\\
&\mathcal{H}=Q\dot{\phi}-\mathcal{L}=\frac{1}{2C}Q^2+\frac{1}{2L}\phi^2,
\end{split}
\end{equation}
with Poisson bracket $\{\phi,Q\}=1$. Note that $Q$ represents the physical charge on the island associated to the node flux $\phi$.
The validity of the Lagrangian method explained above is not restricted to the parallel LC circuit: it holds for any network made of inductive and capacitive elements, since it correctly describes each term we would usually write in the Kirchhoff's laws. However, it does not consider the last category of passive circuit elements, i.e. resistors, since it would include a cumbersome and unwieldy expression to describe the energy dissipated by the resistor at time $t$. How to include resistors in the Hamiltonian formalism is the subject of Sec.~\ref{sec:model}.

\begin{wrapfigure}{r}{0.45\textwidth}
  \begin{center}
    \includegraphics[scale=0.4]{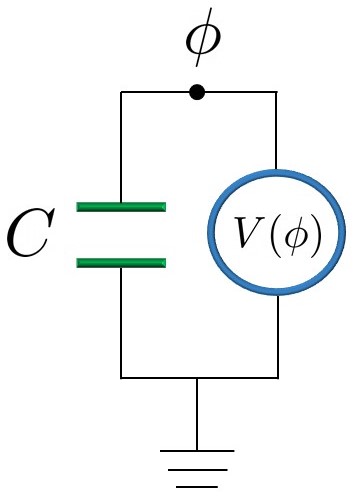}
  \end{center}
  \caption{Effective diagram to describe a Josephson junction, or, more generally, a superconducting qubit. The ``potential energy'' $V(\phi)$ is the inductive energy of the junction, namely $V(\phi)=-E_j\cos\varphi=-E_j\cos\left(2\pi\frac{\phi}{\phi_0}\right)$, where $E_j$ is an energy constant and $\phi_0$ is the magnetic flux quantum.}
  \label{fig:josephson}
\end{wrapfigure}

Now that we have identified the Poisson brackets of the degrees of freedom of the system and written its Hamiltonian, we are ready to quantize the circuit. However, first of all we should explain the meaning of such a procedure for electromagnetic circuits; indeed, it is not trivial to understand when and why a mesoscopic circuit behaves quantum-mechanically, i.e. when all of its constituents are not well described anymore by a classical approximation. 
The requirements for the quantum regime basically reads \cite{Girvin2014}: the circuit must be made of superconducting materials, and must be kept at a temperature (usually around 20 mK) such that the thermal excitations are not comparable with the superconducting energy gap necessary to break a Cooper's pair. Moreover, additional excitations such as plasmons are usually of a frequency way higher than the relevant frequencies of the circuit (typically of a few GHz). Therefore, the electrons in the circuit are bulked in a superconducting collective state that cannot be broken by any excitation, and a quantum-mechanical description is meaningful and necessary. We refer the reader to Refs.~\cite{Vool2017,Devoret2004,Girvin2014} for an extensive discussion on this topic.

Let us go back to the Hamiltonian of the parallel LC circuit in Eq.~\eqref{eqn:hamiltonianLC}. If the conditions discussed above are fulfilled, we can quantize the macroscopic flux and charge of the circuit and write:
\begin{equation}
\label{eqn:quantization}
\begin{split}
&\phi\rightarrow \hat{\phi},\qquad Q\rightarrow \hat{Q},\qquad\{\phi,Q\}=1\rightarrow [\hat{\phi},\hat{Q}]=i\hbar.
\end{split}
\end{equation}
From now on, we drop the hat sign over the operators.
The quantum circuit Hamiltonian is now the Hamiltonian of a quantum harmonic oscillator with frequency $\omega=\frac{1}{\sqrt{LC}}$, which is the resonant frequency of the LC circuit. We can define the standard creation and annihilation operators as:
\begin{equation}
\label{eqn:creationAnnihilation}
\begin{split}
&a=\frac{i}{\sqrt{2C\hbar\omega}}Q+\frac{1}{\sqrt{2L\hbar\omega}}\phi,\qquad a^\dagger=\frac{-i}{\sqrt{2C\hbar\omega}}Q+\frac{1}{\sqrt{2L\hbar\omega}}\phi,
\end{split}
\end{equation}
and the Hamiltonian reads
\begin{equation}
\label{eqn:quantizedLCham}
\mathcal{H}=\hbar \omega\left(a^\dagger a+\frac{1}{2}\right).
\end{equation}
A necessary circuit element to create superconducting qubits is the Josephson junction, i.e. two superconducting electrodes connected by a thin insulating barrier \cite{Josephson1962}. The first ever superconducting qubit, the so-called \textit{charge qubit}, essentially consisted  of a Josephson junction shunted by a voltage source \cite{Bouchiat1998}. As depicted in Fig.~\ref{fig:josephson}, the Josephson junction can be described by a capacitor in parallel with a non-linear inductance, giving rise to a ``potential energy'' of the form $U_L=-E_j\cos\varphi$, where $E_j$ is an energy constant depending on the junction and $\varphi$ is a variable proportional to the flux $\phi$ through it. By employing this non-linear effect, it is possible to isolate the two lowest quantum states of the junction Hamiltonian and to fabricate a qubit (see Refs. \cite{Makhlin2001,Girvin2014,Wendin2016} for a review).
For the purpose of Hamiltonian analysis of a quantum circuit, we describe superconducting qubits, or Josephson junctions, as circuit elements made of a capacitor in parallel with a source of (inductive) potential energy $V(\phi)$, as shown in Fig.~\ref{fig:josephson}.

\mar{
\section{Fourier Transform in quantum mechanics versus signal analysis}
\label{sec:sign}
We give here some clarifications on the opposite sign choices in the definition of Fourier transform in quantum mechanics and signal analysis.

Consider for example a quantum harmonic oscillator with frequency $\omega_{0}$ and Hamiltonian $H = \hbar \omega_{0}(a^{\dag}a + 1/2)$. The Heisenberg equations of motion yield $a(t) = e^{-i \omega_{0} t} a(0)$ and $a^{\dag}(t) = e^{i \omega_{0} t} a^{\dag}(0)$, and the position operator is
\begin{equation}
x(t) = \sqrt{\frac{\hbar}{2m\omega_{0}}}a(0)e^{-i\omega_{0}t} + \sqrt{\frac{\hbar}{2m\omega_{0}}}a^{\dag}(0)e^{i\omega_{0}t},
\end{equation}
where m is the mass. 
We can see that both components appear, one corresponding to the annihilation operator and rotating clockwise in the complex plane, and the other corresponding to the creation operator and rotating counterclockwise. None of them is ``more fundamental'' and since they are non-Hermitian operators they are not observables. The observable is the Hermitian position operator $x(t)$. However, it is convenient to write the equations of motion in terms of the annihilation operator $a$, to avoid carrying along the dagger sign. Thus, for the Fourier transform we would like to have a convention that attributes the Fourier component at frequency $\omega_{0}$ to the annihilation operator. The definition Eq. (\ref{eq:Fourier}) clearly achieves this, since $a (\omega ) = 2\pi a(0) \delta (\omega -\omega_{0} )$, but we stress that in the Hermitian $x(t)$ both Fourier components are present,
\begin{equation}
x(\omega ) = \sqrt{\frac{\pi h}{m\omega_{0}}} a(0) \delta (\omega - \omega_{0} ) + \sqrt{\frac{\pi h}{m\omega_{0}}} a^{\dag}(0) \delta (\omega + \omega_{0} ).
\end{equation}
Now, let us examine the conventions from signal analysis and electrical-circuits engineering. 
Consider a generic harmonic signal $x(t) = X \cos (\omega_{0} t + \phi )$, which can be also written as
\begin{equation}
x(t) = \frac{X}{2}e^{-i \phi} e^{-i \omega_{0} t} + \frac{X}{2}e^{i\phi} e^{i \omega_{0} t},
\end{equation}
with Fourier transform
\begin{equation}
x(\omega ) = \pi X e^{-i\phi}\delta (\omega-\omega_{0} ) + \pi X e^{i\phi}  \delta (\omega + \omega_{0}) .
\end{equation}
Again it is clear that both frequency components are present but we have to make a choice. This choice in electrical engineering is motivated by the representation of complex numbers as phasors. Phasors can be summed by the usual rules of voltage addition, thus, several signals at the same frequency can be analysed in a time-independent framework.
For the signal above, the standard definition of its phasor is $Xe^{i\phi}$, with the property that angles are measured counterclockwise in the complex plane, and the real signal $x(t)$ is then obtained as $x(t) =  \mathrm{Re}[Xe^{i\phi } e^{i\omega_{0} t}]$. However, we can see that there is nothing fundamental in this choice. Equally well, we could designate $Xe^{-i\phi}$ and follow its evolution, and finally write $x(t) =  \mathrm{Re}[Xe^{-i\phi } e^{-i\omega_{0} t}]$.
Doing so however changes the standard textbook definition of the complex ac impedances and admittances for standard components such as inductors and capacitors, as used in electrical engineering. If we wish to recover these definitions, we simply have to replace $\omega \rightarrow -\omega$ in all the results of the paper.
}

\section{Linear response function and generalized impedance}
\label{sec:linearResp}
Let us start with the definition of complex impedance Eq.~\eqref{eqn:impedDef}:
\begin{equation}
\label{eqn:fourierRel}
V(\omega)=Z(\omega)I(\omega).
\end{equation}
Provided it is well-defined, we can take the Fourier transform of the above equation and obtain:
\begin{equation}
\label{eqn:directRel}
V(t)=\frac{1}{2\pi}\int_{-\infty}^\infty d\omega\, Z(\omega)e^{-i\omega t}\int_{-\infty}^\infty dt'\,e^{i\omega t'}I(t'),
\end{equation}
and assuming that the integration order can be switched,
\begin{equation}
\label{eqn:directRelBis}
V(t)=\int_{-\infty}^\infty dt'\,Z(t-t')I(t'),
\end{equation}
which coincides with Eq.~\eqref{eqn:linearResponseImpedance}. Both the above assumptions, however, are quite strong, and for instance the impedance function of a parallel LC circuit does not fulfil them. We clearly understand that this is not a physical problem, but a mathematical issue due to the divergence of the impedance function in a zero-measure set. We can therefore try to avoid this pathological point of the frequency range	 by slightly changing the definition of impedance.

The trick consists in introducing a real parameter $\eta$ in Eq.~\eqref{eqn:directRelBis}, and then taking its value to zero so that it cannot change any physical result\footnote{Given that we are working in the time-domain, which can be considered as the ``really physical'' one.}, as done in Eq.~\eqref{eqn:linearResponseImpedanceGen}. The negative exponential does not cause any trouble because $Z(t)$ vanishes for negative times \cite{Jackson}. At this point, we can follow the inverse path that led us to Eq.~\eqref{eqn:directRelBis}, and write:
\mar{
\begin{equation}
\begin{split}
  V(\omega)=&\lim_{\eta\rightarrow 0^+}\int_{-\infty}^\infty dt e^{i\omega t}\,\int_{-\infty}^\infty dt'\,Z(t-t')e^{-\eta (t-t')}I(t').\\
  =&\int_{-\infty}^\infty dt' e^{i\omega t'}I(t')\lim_{\eta\rightarrow 0^+} \int_{-\infty}^\infty dt\, e^{i\omega (t-t')} Z(t-t')e^{-\eta (t-t')} \\
  =& \int_{-\infty}^\infty dt' e^{i\omega t'}I(t') Z^g(\omega)=   Z^g(\omega) I(\omega).
\end{split}
\end{equation}
}
Thanks to the negative exponential, the Fourier transform of $Z(t)$ is now well-defined and the integrals can be interchanged. Hence, we have finally obtained the desired relation Eq.~\eqref{eqn:fourierRel} with the new generalized impedance defined in Eq.~\eqref{eqn:generalizedImpedance}.

\section{Derivation of the qubit-bath interaction}
\label{sec:derivation}
In this appendix we present all the derivations of the circuit Hamiltonians presented in Sec.~\ref{sec:coupling}, based on the method introduced in Ref.~\cite{Parra-Rodriguez2018}. We do not reproduce here the circuit diagrams, and we refer the interested reader to Sec.~\ref{sec:coupling}.

\subsection{Single qubit}
\label{sec:appendixSingle}
In what follows we  discuss the coupling of one qubit to a resistor via a capacitor with capacitance $C_g$, as shown in Fig.~\ref{fig:oneQubit}. We describe the resistor using the Caldeira-Leggett model presented in Sec.~\ref{sec:Devoret}, thus considering as internal degrees of freedom the voltage differences across each LC circuit of the Foster's first form, associated to the fluxes $\phi_j$ with $j=1,2,\ldots$. From now on, we use greek letters to refer to such internal degrees of freedom, therefore we  write $\phi_\alpha$ instead of $\phi_j$.

The Lagrangian of the circuit in Fig.~\ref{fig:oneQubit} is readily written:
\begin{equation}
\label{eqn:lagrangianSingleQ}
\begin{split}
\mathcal{L}&=\frac{C_A}{2}\dot{\phi}_A^2-V(\phi_A)+\sum_{\alpha=1}^\infty \left[\frac{C_\alpha}{2}\dot{\phi}_\alpha^2-\frac{1}{2L_\alpha}\phi_\alpha^2\right]+\frac{C_g}{2}\left(\dot{\phi}_A-\sum_{\alpha=1}^\infty \dot{\phi}_\alpha\right)^2\\
&=\frac{1}{2}\dot{\vecP}^T C\dot{\vecP}-\frac{1}{2}\vecP^T L^{-1}\vecP -V(\phi_A),
\end{split}
\end{equation}
where we have introduced the vector $\vecP=(\phi_A,\phi_1,\phi_2,\ldots)^T$ and the capacitance matrix
\begin{equation}
\label{eqn:capMatrixSingleQ}
C=\begin{pmatrix}
C_A+C_g&-C_g\bm{e}_\alpha^T\\
-C_g\bm{e}_\alpha&\mathsf{C}_\alpha+C_g\bm{e}_\alpha\bm{e}_\alpha^T
\end{pmatrix},
\end{equation}
where $\bm{e}_\alpha=(1,1,\ldots)^T$ is an infinite vector filled with ones, and $\mathsf{C}_\alpha=diag(C_1,C_2,\ldots)$ contains the values of the capacitances of the internal degrees of freedom of the resistor. We have also introduced the \textit{inductance matrix}, defined as
\begin{equation}
\label{eqn:inductanceMatrixSingleQ}
L^{-1}=\begin{pmatrix}
0&\bm{0}_\alpha^T\\
\bm{0}_\alpha&\mathsf{L}^{-1}_\alpha
\end{pmatrix},
\end{equation}
where $\mathsf{L}^{-1}_\alpha=diag(L_1^{-1},L_2^{-1},\ldots)$.

Our aim is to apply the Legendre transformation to the above Lagrangian and obtain the Hamiltonian. Moreover, we want such Hamiltonian to be in a readable form, that is to say, we want it to be divided into three parts: the free Hamiltonian of the qubit, the free Hamiltonian of the resistor as a thermal bath, and the interaction Hamiltonian coupling the two elements. This corresponds to writing, in the final Hamiltonian, a capacitance matrix $C^{-1}$ such that the (infinite) $\alpha\times\alpha$ submatrix associated to the degrees of freedom of the resistor is diagonal, and the same for the inductance matrix. In order to do so, we rely on a method first employed by Paladino et al. \cite{Paladino2003}, and very recently improved in Ref.~\cite{Parra-Rodriguez2018}.

The idea consists in searching for a coordinate transformation (a point transformation in the language of Hamiltonian mechanics) such that the new coordinates read $\bm{z}=Z\vecP$, where $Z$ is a symmetric matrix, and the new associated capacitance (inductance) matrix is given by $C_z=Z^{-1}CZ^{-1}$ ($L_z^{-1}=Z^{-1}L^{-1}Z^{-1}$). We choose to parametrize such point transformation as
\begin{equation}
\label{eqn:pointTransformationZ}
Z=\begin{pmatrix}
1&\mathbf{0}_\alpha^T\\
\mathbf{0}_\alpha&M_0^{-1/2}\mathsf{M}_\alpha^{1/2}
\end{pmatrix},
\end{equation}
with $\mathsf{M}_\alpha=\mathsf{C}_\alpha+\xi \bm{e}_\alpha\bm{e}_\alpha^T$, where $\xi$ is a parameter we can freely choose, and $M_0$ is a free constant with the units of capacitance, whose value does not induce any physical effect.
The new capacitance matrix reads
\begin{equation}
\label{eqn:newCapacitanceSingleQ}
C_z=\begin{pmatrix}
C_{\Sigma_A}&-C_g \bm{f}_\alpha^T\\
-C_g\bm{f}_\alpha&M_0\mathds{I}+(C_g-\xi)\bm{f}_\alpha\bm{f}_\alpha^T
\end{pmatrix},
\end{equation}
where for convenience we have introduced the notation $C_{\Sigma_A}=C_A+C_g$, and $\bm{f}_\alpha=M_0^{1/2}\mathsf{M}_\alpha^{-1/2}\bm{e}_\alpha$.

In order to obtain the Hamiltonian as a function of $\bm{z}$ and their conjugate momenta, we now have to find $C_z^{-1}$. This inversion is obtained through Eq.~\eqref{eqn:inversionFormulaSpecific}	 proven in Appendix~\ref{sec:inversion}, in turn derived from a formula presented in Ref.~\cite{Parra-Rodriguez2018}. The inverse matrix reads:
\begin{equation}
\label{eqn:inverseSingleQ}
C_z^{-1}=\begin{pmatrix}
\frac{\alpha}{D}&\frac{\beta}{D}\bm{f}_\alpha^T\\
\frac{\beta}{D}\bm{f}_\alpha&M_0^{-1}\mathds{I}+\frac{\delta}{D}\bm{f}_\alpha\bm{f}_\alpha^T
\end{pmatrix},
\end{equation}
with coefficients
\begin{equation}
\label{eqn:coeffSingleQ}
\begin{split}
\alpha=&M_0+(C_g-\xi)\abs{\bm{f}_\alpha}^2,\\
\beta=&C_g,\\
\delta=&(C_g^2-C_{\Sigma_A}(C_g-\xi))/M_0,\\
D=&M_0C_{\Sigma_A}+[(C_g-\xi)C_{\Sigma_A}-C_g^2]\abs{\bm{f}_\alpha}^2.\\
\end{split}
\end{equation}
In order to remove any interaction between the modes of the resistor, from Eq.~\eqref{eqn:inverseSingleQ} it is clear that we need to set $\delta=0$, i.e. we choose the parameter $\xi=C_g-C_g^2/C_{\Sigma_A}$, and finally we obtain
\begin{equation}
\label{eqn:inverseSingleQbis}
C_z^{-1}=\begin{pmatrix}
C_{\Sigma_A}^{-1}(1+\abs{\bm{f}_\alpha}^2C_g^2C_{\Sigma_A}^{-1} M_0^{-1})\quad&C_g C_{\Sigma_A}^{-1} M_0^{-1}\bm{f}_\alpha^T\\
C_g C_{\Sigma_A}^{-1} M_0^{-1}\bm{f}_\alpha&M_0^{-1}\mathds{I}
\end{pmatrix}.
\end{equation}

$\abs{\bm{f}_\alpha}^2$ is the square norm of an infinite vector, hence we need to worry about a possible divergence. Fortunately, as proven in Ref. \cite{Parra-Rodriguez2018} this norm always converges, since by using the Sherman-Morrison inversion formula we get
\begin{equation}
\label{eqn:shermanMorrison}
\mathsf{M}_\alpha^{-1}=\mathsf{C}_\alpha^{-1}-\xi\frac{\mathsf{C}_\alpha^{-1}\bm{e}_\alpha\bm{e}_\alpha^T\mathsf{C}_\alpha^{-1}}{1+\xi\bm{e}_\alpha^T\mathsf{C}_\alpha^{-1}\bm{e}_\alpha},
\end{equation} 
therefore
\begin{equation}
\begin{split}
\abs{\bm{f}_\alpha}^2&=M_0\bm{e}_\alpha^T\mathsf{M}_\alpha^{-1}\bm{e}_\alpha=\frac{M_0\bm{e}_\alpha^T\mathsf{C}_\alpha^{-1}\bm{e}_\alpha}{1+\xi\bm{e}_\alpha^T\mathsf{C}_\alpha^{-1}\bm{e}_\alpha},
\end{split}
\end{equation}
which is finite even in the pathological case $\bm{e}_\alpha^T\mathsf{C}_\alpha^{-1}\bm{e}_\alpha\rightarrow\infty$. Anyway, if we choose the values of $\mathsf{C}_\alpha$ to represent a resistor as in Eq.~\eqref{eqn:valueDevoretComega} and Eq.~\eqref{eqn:impedanceOhmic}, this quantity remains finite as well.

If $\bm{p}$ are the momenta associated to the transformed coordinates $\bm{z}$, the Hamiltonian is now given by
\begin{equation}
\label{eqn:HamiltonianStep2}
\mathcal{H}=\frac{1}{2}\bm{p}^T C_z^{-1}\bm{p}+\frac{1}{2}\bm{z}^T L_z^{-1}\bm{z} +V(\phi_A).
\end{equation}
We have finally managed to decouple the internal modes of the resistor in the capacitance matrix. Unfortunately, this is not true for the inductance matrix, which now reads
\begin{equation}
\label{eqn:newInductanceSingleQ}
L_z^{-1}=\begin{pmatrix}
0&\bm{0}_\alpha^T\\
\bm{0}_\alpha&M_0\mathsf{M}_\alpha^{-1/2}\mathsf{L}_\alpha^{-1}\mathsf{M}_\alpha^{-1/2}
\end{pmatrix}.
\end{equation}
Therefore, in order to get the Hamiltonian into the desired form, we need to find the unitary transformation that diagonalizes $L_z^{-1}$, by calculating the eigenvalues and eigenvectors of Eq.~\eqref{eqn:newInductanceSingleQ}. This requires finding $\mathsf{M}_\alpha^{-1/2}$, and none of these tasks can lead to an explicit formula for a generic system. After knowing the value of all the relevant circuit parameters, we would most likely need to cut off a certain number of modes in order to deal with finite matrices, and then to perform the diagonalization numerically. 

\mar{It is worth pointing out that a closed-form expression for the Hamiltonian can, in general, be found by following a different procedure, which relies on a description of the dissipative part of the network as a generic transmission line. This leads to solving a boundary condition differential equation singular value problem. The essential difference lies in the fact that the internal modes of the resistive element are there treated as a continuum, in contrast with the Caldeira-Leggett method we employ in this paper, where the modes are discrete and obtained through a Foster's decomposition. This alternative approach has been differently introduced and developed in Refs.~\cite{Bamba2014,Malekakhlagh2016b,Parra-Rodriguez2018,Parra-Rodriguez2019}, and an extensive review can be found in the Ph.D. thesis by Adrián Parra-Rodriguez \cite{Parra-Rodriguez2021}. As already said, in this paper we do not discuss this approach, and we just focus on the Caldeira-Leggett method, so as to recover the standard description of a thermal bath as a collection of infinite, discrete bosonic modes. We will later see that a closed-form expression for the Hamiltonian can be obtained also in the weak coupling limit of the Caldeira-Leggett model.}

The suitable unitary transformation $U$ takes the form
\begin{equation}
\label{eqn:unitaryTransf}
U=\begin{pmatrix}
1&\bm{0}_\alpha^T\\
\bm{0}_\alpha&\mathsf{U}_\alpha
\end{pmatrix},
\end{equation}
and the variables  transform according to $\bm{z}'=U\bm{z}$, $\bm{p}'=U\bm{p}$. $U$ is a canonical transformation, i.e. it preserves the Poisson brackets. $L_z^{-1}$ is diagonalized through ${L_z'}^{-1}=U{L_z}^{-1}U^T$, while the capacitance matrix is rotated according to ${C_z'}^{-1}=U{C_z}^{-1}U^T$, which only affects the coupling vector $\bm{f}_\alpha$, transformed into $\bm{f}_\alpha'=\mathsf{U}_\alpha\bm{f}_\alpha$.

Finally, the Hamiltonian $\mathcal{H}=\frac{1}{2}\bm{p}'{C_z'}^{-1}\bm{p}'+\frac{1}{2}\bm{z}'{L_z'}^{-1}\bm{z}'+V(\phi_A)$ can be rewritten as\footnote{Note that in Eq.~\eqref{eqn:HamiltonianSingleQstrong} we are using $\alpha$ both to indicate the $\alpha$th internal mode of the resistor and as an index of the matrices or vectors thereof.}
\begin{equation}
\label{eqn:HamiltonianSingleQstrong}
\begin{split}
\mathcal{H}&=\frac{({C_z'}^{-1})_{11}}{2}p_A^2+V(\phi_A)+\frac{C_g}{M_0C_{\Sigma_A}}\sum_{\alpha=1}^\infty  f'_\alpha p_Ap_\alpha' +\sum_{\alpha=1}^\infty \frac{1}{2}\left[\frac{p_\alpha'^2}{M_0}+(L_z'^{-1})_{\alpha\alpha}z_\alpha'^2\right].\\
\end{split}
\end{equation}
We observe that, by capacitively coupling a qubit to a resistor, the interaction Hamiltonian displays a qubit-bath coupling through the qubit momentum $p_A$ with the units of charge (through $\sigma^y$ in the case of a transmon qubit), in accordance with the discussion on the capacitive coupling in Sec.~\ref{sec:capCoupling}. Therefore, the capacitance matrix inversion and the search for a diagonalizing unitary $U$ are necessary only to obtain the weight of the coupling to each bath mode, described by the vector $\bm{f}_\alpha'$ in Eq.~\eqref{eqn:HamiltonianSingleQstrong}, which also defines the spectral density. We can absorb the constant $M_0$ by transforming the bath quadratures into annihilation and creation operators of each mode (respectively $a_\alpha$ and $a_\alpha^\dagger$) \cite{MandelWolf}:
\begin{equation}
\label{eqn:HamiltonianSingleQstrongBis}
\begin{split}
\mathcal{H}=&\frac{({C_z'}^{-1})_{11}}{2}p_A^2+V(\phi_A)+i\frac{C_g}{C_{\Sigma_A}}\sum_{\alpha=1}^\infty  \sqrt{\frac{\hbar\omega_\alpha}{2M_0}}f'_\alpha p_A (a_\alpha^\dagger-a_\alpha)+\sum_{\alpha=1}^\infty \hbar\omega_\alpha\left(a_\alpha^\dagger a_\alpha+\frac{1}{2}\right).\\
\end{split}
\end{equation}
$\omega_\alpha=\sqrt{(L_z^{-1})_{\alpha\alpha}/M_0}$ are the frequencies of each resistor mode, forming a continuum, while the constant $M_0$ in the interaction Hamiltonian is going to be absorbed by the coefficients $f_\alpha'$, according to their definition in Eq.~\eqref{eqn:newCapacitanceSingleQ}. Following the convention of quantum optics, we  drop the infinite constant term in the resistor Hamiltonian. Finally, note that the qubit Hamiltonian $\mathcal{H}_q=\frac{1}{2}({C_z'}^{-1})_{11}p_A+V(\phi_A)$ has been renormalized by the interaction with the resistor, since in general $({C_z'}^{-1})_{11}\neq C_A^{-1}$. This is an important effect of circuit QED, which may be negligible in the weak coupling limit, but may also display non-trivial effects in broader scenarios \cite{Paladino2003}.

\subsubsection{Weak coupling limit}
\label{sec:weakcoupling}
Despite a closed-form expression for the Hamiltonian is in general not available, i.e. we cannot know a priori all the coefficients of Eq.~\eqref{eqn:HamiltonianSingleQstrong} as functions of the circuit parameters, in the weak coupling limit we are able to recover a closed-form expression.
Defining the weak coupling limit in the presence of a resistor is less trivial than in Sec.~\ref{sec:capCoupling}. Indeed, while there we could set $C_g\ll C_A,C_B$, the internal degrees of freedom of the resistor are related to an infinite collection of capacitances stored in the diagonal matrix $\mathsf{C}_\alpha$, expressed in Eq.~\eqref{eqn:valueDevoretComega}. We set the weak coupling limit as $C_g\ll C_A$, $C_g\ll (\mathsf{C}_\alpha)_{jj}$. If this is the case\footnote{Taking the mathematical limit $\Delta \omega\rightarrow 0$, we have $(\mathsf{C}_\alpha)_{jj}\rightarrow\infty$ and the condition is satisfied. In a model where $\Delta \omega$ is very small but finite, one has to compare its value with the resistance $R$ and the capacitance $C_g$ to check whether the weak coupling limit is satisfied.}, we can neglect the effects of the order of $C_g/(\mathsf{C}_\alpha)_{jj}$ in the point transformation and in the renormalization of the internal modes of the resistor, and the qubit-bath coupling is weak compared to the energy of the qubit (with a perturbation parameter of the order of $O(C_g/C_A)$). In this case, recalling Eq.~\eqref{eqn:shermanMorrison} with $\xi=C_g-C_g^2/C_{\Sigma_A}$, at the zeroth order we have  $\mathsf{M}_\alpha^{-1}\approx \mathsf{C}_\alpha^{-1}$, and we obtain the final Hamiltonian:
\begin{equation}
\label{eqn:weakHamSingleQ}
\begin{split}
\mathcal{H}\approx&\frac{p_A^2}{2C_{\Sigma_A}}+V(\phi_A)+i\frac{C_g}{C_{\Sigma_A}}\sum_{\alpha=1}^\infty  \sqrt{\frac{\hbar\omega_\alpha \Re[Z(\omega_\alpha)]\Delta\omega}{\pi}}p_A (a_\alpha^\dagger-a_\alpha) +\sum_{\alpha=1}^\infty \hbar\omega_\alpha a_\alpha^\dagger a_\alpha.\\
\end{split}
\end{equation}
Note that requiring $C_g\ll (\mathsf{C}_\alpha)_{jj}$ is sufficient to obtain a closed-form expression for the interaction Hamiltonian, while the additional condition $C_g\ll C_A$ introduces the standard weak coupling limit in the language of open quantum systems, for which the system-bath interaction is a perturbation that can be treated through the Born-Markov approximation, leading to the master equation~\eqref{eqn:masterEqSingleQ}.

\subsection{Two qubits, common bath}
\label{sec:appendixTwoQComm}
The two-qubit scenario is similar to the single qubit one presented in Sec.~\ref{sec:singleQ}, although there are some subtleties that are worth discussing.
Let us start with the case where the two qubits, named $A$ and $B$, are capacitively coupled to the same resistor, but not directly coupled together, as depicted in Fig.~\ref{fig:twoQ}. This corresponds to the well-known ``common bath'' scenario in the language of open quantum systems. For simplicity, we consider the balanced case in which the two coupling capacitors have equal capacitances; the derivation of the Hamiltonian in the unbalanced case can be obtained by following the same method we discuss here. 

The circuit Lagrangian is:
\begin{equation}
\label{eqn:Lagrangian2Q}
\begin{split}
\mathcal{L}=&\frac{C_A}{2}\dot{\phi}_A^2-V(\phi_A)+\frac{C_B}{2}\dot{\phi}_B^2-V(\phi_B)+\sum_{\alpha=1}^\infty	\left[\frac{C_\alpha}{2}\dot{\phi}_\alpha^2-\frac{1}{2L_\alpha}\phi_\alpha^2\right]\\
&+\frac{C_g}{2}\left(\dot{\phi}_A-\sum_{\alpha=1}^\infty\dot{\phi}_\alpha\right)^2+\frac{C_g}{2}\left(\dot{\phi}_B-\sum_{\alpha=1}^\infty\dot{\phi}_\alpha\right)^2\\
=&\frac{1}{2}\dot{\vecP}^T C\dot{\vecP}-\frac{1}{2}\vecP^T L^{-1}\vecP -\sum_{j=A,B}V(\phi_J),
\end{split}
\end{equation}
where we have used the same notation of the previous section in Eq.~\eqref{eqn:lagrangianSingleQ}, with $\vecP=(\phi_A,\phi_B,\phi_1,\phi_2,\ldots)^T$. The capacitance matrix reads:
\begin{equation}
\label{eqn:capacitanceTwoQ}
C=\begin{pmatrix}
\mathsf{S}&-C_g\bm{a}\bm{e}_\alpha^T\\
-C_g\bm{e}_\alpha\bm{a}^T&\mathsf{C}_\alpha+2C_g\bm{e}_\alpha\bm{e}_\alpha^T
\end{pmatrix},
\end{equation}
where $\bm{a}=(1,1)^T$, $\bm{e}_\alpha=(1,1,\ldots)^T$ and $\mathsf{S}$ is the system matrix, describing the qubits:
\begin{equation}
\label{eqn:systemMatrix}
\mathsf{S}=\begin{pmatrix}
C_A+C_g&0\\
0&C_B+C_g
\end{pmatrix}=\begin{pmatrix}
C_{\Sigma_A}&0\\
0&C_{\Sigma_B}
\end{pmatrix}.
\end{equation}
Note that any possible imbalance between the qubit-bath couplings, e.g. when the coupling capacitors have different capacitances, is reflected in the elements of the vector $\bm{a}$, which in this case would not be equal anymore.
The inductance matrix $L^{-1}$ is the same as in Eq.~\eqref{eqn:inductanceMatrixSingleQ}, with an additional zero row and column because of the second qubit.

Following the path of the previous section, we introduce new coordinates through the point transformation Eq.~\eqref{eqn:pointTransformationZ} (once again with additional zero row and column), and obtain the new capacitance matrix given by
\begin{equation}
\label{eqn:newCapacitanceTwoQ}
C_z=\begin{pmatrix}
\mathsf{S}&-C_g \bm{a}\bm{f}_\alpha^T\\
-C_g\bm{f}_\alpha\bm{a}^T&M_0\mathds{I}+(2C_g-\xi)\bm{f}_\alpha\bm{f}_\alpha^T
\end{pmatrix},
\end{equation}
with $\bm{f}_\alpha=M_0^{1/2}\mathsf{M}_\alpha^{-1/2}\bm{e}_\alpha$, $M_0$ is an irrelevant constant with the units of capacitance and $\mathsf{M}_\alpha$ is defined in Eq.~\eqref{eqn:pointTransformationZ}.

As proven in Appendix~\ref{sec:inversion}, the inverse of Eq.~\eqref{eqn:newCapacitanceTwoQ} reads
\begin{equation}
\label{eqn:inverseTwoQ}
C_z^{-1}=\begin{pmatrix}
\mathsf{S}^{-1}+C_g^2\abs{\bm{f}_\alpha}^2D^{-1}\mathsf{S}^{-1}\bm{a}\bm{a}^T\mathsf{S}^{-1}&C_g D^{-1}\mathsf{S}^{-1}\bm{a}\bm{f}_\alpha^T\\
C_g D^{-1}\bm{f}_\alpha\bm{a}^T\mathsf{S}^{-1}&M_0^{-1}\mathds{I}+[\bm{a}^T\mathsf{S}^{-1}\bm{a}(C_g^2-(2C_g-\xi))/M_0]D^{-1}\bm{f}_\alpha\bm{f}_\alpha^T
\end{pmatrix},
\end{equation}
with $D=M_0+\abs{\bm{f}_\alpha}^2\left[(2C_g-\xi)-C_g^2\bm{a}^T\mathsf{S}^{-1}\bm{a}\right]$.
In order to remove the capacitive interaction between different modes, we choose $\xi=2C_g-\Tr\mathsf{S}^{-1}C_g^2$, so that the submatrix describing the resistor modes (block $(2,2)$ of Eq.~\eqref{eqn:inverseTwoQ}) is proportional to the identity matrix and $D=M_0$. Therefore, the inverse capacitance matrix is now given by
\begin{equation}
\label{eqn:inverseTwoQbis}
C_z^{-1}=\begin{pmatrix}
\mathsf{S}^{-1}+C_g^2 M_0^{-1}\abs{\bm{f}_\alpha}^2\mathsf{N}&C_g M_0^{-1}\bm{s}\bm{f}_\alpha^T\\
C_g M_0^{-1}\bm{f}_\alpha\bm{s}^T&M_0^{-1}\mathds{I}
\end{pmatrix},
\end{equation}
where $\bm{s}=\mathsf{S}^{-1}\bm{a}=(C_{\Sigma_A}^{-1},C_{\Sigma_B}^{-1})^T$, and
\begin{equation}
\label{eqn:matrixN}
\mathsf{N}=\mathsf{S}^{-1}\bm{a}\bm{a}^T\mathsf{S}^{-1}=\begin{pmatrix}
C_{\Sigma_A}^{-2}&C_{\Sigma_A}^{-1}C_{\Sigma_B}^{-1}\\
C_{\Sigma_A}^{-1}C_{\Sigma_B}^{-1}&C_{\Sigma_B}^{-2}
\end{pmatrix}.
\end{equation}

We now face the same issue as in the previous section: while there is no capacitive interaction between the resistor modes, the inductive matrix is not diagonal. We need to numerically find the suitable canonical transformation Eq.~\eqref{eqn:unitaryTransf} in order to completely decouple the modes of the bath, and find the spectral density of their interaction with the qubit. Anyway, although we are not able to find a priori a closed-form expression for the Hamiltonian, we can already observe that the qubits are not only coupled to the same bath, but a direct qubit-qubit interaction term appears in the Hamiltonian, proportional to $\sigma^y_A\sigma^y_B$ in the case of two transmon qubits. This is the effect of the non-diagonal matrix $\mathsf{N}$ appearing in the block $(1,1)$ of Eq.~\eqref{eqn:inverseTwoQbis}, and is in accordance  with the result found when introducing the capacitive coupling in Sec.~\ref{sec:capCoupling}.
As in the case of a single qubit, we are still able to find a closed-form expression for the Hamiltonian in the weak coupling limit.

\subsubsection{Weak coupling limit}
The weak coupling limit applies as in the case of a single qubit, as explained in Sec.~\ref{sec:weakcoupling}: we assume $C_g\ll (\mathsf{C}_\alpha)_{jj}$ to get a closed-form expression for the interaction Hamiltonian, and $C_g\ll C_A,C_B$ to remove any long-range coupling between the qubits and to recover the weak coupling limit of the system-bath interaction. With these prescriptions, if $\bm{p}$ are the momenta associated to the coordinates $\bm{z}$, the circuit Hamiltonian now reads:
\begin{equation}
\label{eqn:weakCouplTwoQ}
\begin{split}
\mathcal{H}&\approx\frac{1}{2C_{\Sigma_A}}p_A^2+V(\phi_A)+\frac{1}{2C_{\Sigma_B}}p_B^2+V(\phi_B)+\sum_{\alpha=1}^\infty\hbar\omega_\alpha a_\alpha^\dagger a_\alpha\\
&+i C_g\sum_{\alpha=1}^\infty\sqrt{\frac{\hbar\omega_\alpha\Re[Z(\omega_\alpha)]\Delta\omega}{\pi}}\left(\frac{p_A}{C_{\Sigma_A}}+\frac{p_B}{C_{\Sigma_B}}\right)(a_\alpha^\dagger-a_\alpha),
\end{split}
\end{equation}
where following the discussion for the single qubit scenario we have introduced the creation and annihiliation operators of each bath mode (respectively $a_\alpha^\dagger$ and $a_\alpha$).

Note that, because of the assumption $C_g\ll C_A,C_B$, any direct interaction between the two qubits is removed. This is due to the fact that in the block $(1,1)$ of Eq.~\eqref{eqn:inverseTwoQbis} the matrix $\mathsf{N}$, inducing the $\sigma_A^y\sigma_B^y$ interaction, appears in a term of the order of $O((C_g/C_{A,B})^2)$, while we are only keeping terms of the first order. This mimics the behaviour of the capacitive coupling in chains of LC circuits discussed in Sec.~\ref{sec:capCoupling}. How can we re-introduce a direct qubit-qubit coupling without dropping the weak coupling approximation? The circuit scheme presented in Fig.~\ref{fig:twoQdirect} does the job: an additional coupling capacitor (weak or strong) directly connecting the qubits modifies the off-diagonal terms of the system matrix $\mathsf{S}$ in Eq.~\eqref{eqn:systemMatrix}. Since the inverse circuit matrix still has the form of Eq.~\eqref{eqn:inverseTwoQ}, we obtain the required interaction. Note that, in this scenario, even if the direct coupling between the qubit $B$ and bath were switched off (i.e. $\bm{a}=(1,0)^T$), we would still obtain an interaction between them in the circuit Hamiltonian, through the vector $\bm{s}=\mathsf{S}^{-1}\bm{a}=(\mathsf{S}^{-1}_{11},\mathsf{S}^{-1}_{12})^T$. A weak qubit-qubit coupling capacitor ($C_c\ll C_{A,B}$) would eliminate this direct interaction, since $\mathsf{S}^{-1}_{12}$ is of the first order in $C_c$.

\subsection{Two qubits, separate baths}
\label{sec:appendixTwoQSep}
Obtaining the general Hamiltonian in the separate baths scenario for a strong coupling is a cumbersome task, since it requires considering two sets of infinite internal variables of the resistors, and finding the inverse of a Lagrangian as a function of all these variables. This would lead to the presence of a direct interaction between all the elements of the circuit (and between the two sets of internal variables as well), greatly complicating the problem. We do not address the general case here, although a solution can still be obtained, by following the path presented in the previous sections and using the tools of Appendix~\ref{sec:inversion} and Ref.~\cite{Parra-Rodriguez2018}. Anyway, we can still find a closed-form expression for the Hamiltonian under the weak coupling approximation: considering the circuit in Fig.~\ref{fig:separateBaths}, the low capacitances $C_g$ and $C_c$ remove all the ``long range'' interactions, in analogy with the case of Sec.~\ref{sec:capCoupling}, and we can obtain the desired Hamiltonian that is presented in Sec.~\ref{sec:separate}. Note that, unlike in the case of a direct coupling and a common bath, in this scenario we must consider a weak direct qubit-qubit coupling as well, otherwise the qubit $B$, for instance, may be directly coupled to the bath on the left in the interaction Hamiltonian. This coupling would be mediated by the vector $\bm{s}$ as in Eq.~\eqref{eqn:inverseTwoQbis}, and it would vanish in the limit $C_c\ll C_{A,B}$, as explained at the end of the weak coupling discussion in Appendix~\ref{sec:appendixTwoQComm}.

\section{Inversion of infinite matrices}
\label{sec:inversion}
The inversion of infinite matrices is an indispensable procedure to find the Hamiltonian of systems of qubits capacitively coupled to resistors, addressed in Sec.~\ref{sec:coupling}. Here we present some useful formula to accomplish the task.

First of all, let us introduce a formula originally obtained in Ref.~\cite{Parra-Rodriguez2018}, valid for single-port impedances, that is, for schemes in which all the elements of the system are coupled to the same point of the resistor. In this case, the capacitance matrix appearing in the Langrangian of the overall system is always of the form:
\begin{equation}
\label{eqn:singlePortMatrix}
\begin{pmatrix}
\mathsf{A}_1&\mathsf{D}\\
\mathsf{D}^\dagger&\mathsf{A}_2
\end{pmatrix},
\end{equation}
where $\mathsf{A}_1$ and $\mathsf{A}_2$ are invertible matrices and $\mathsf{D}$ is a rank-one matrix. The following inversion formula holds \cite{Parra-Rodriguez2018}:
\begin{equation}
\label{eqn:inversionFormula}
\begin{split}
\begin{pmatrix}
\mathsf{A}_1&\mathsf{D}\\
\mathsf{D}^\dagger&\mathsf{A}_2
\end{pmatrix}^{-1}=&\begin{pmatrix}
\mathsf{A}_1^{-1}&0\\
0&\mathsf{A}_2^{-1}
\end{pmatrix}\\
&+\frac{1}{1-\Tr(\mathsf{D}\mathsf{A}_2^{-1}\mathsf{D}^\dagger\mathsf{A}_1^{-1})}\begin{pmatrix}
\mathsf{A}_1^{-1}\mathsf{D}\mathsf{A}_2^{-1}\mathsf{D}^\dagger\mathsf{A}_1^{-1}&-\mathsf{A}_1^{-1}\mathsf{D}\mathsf{A}_2^{-1}\\
-\mathsf{A}_2^{-1}\mathsf{D}^\dagger\mathsf{A}_1^{-1}&\mathsf{A}_2^{-1}\mathsf{D}^\dagger\mathsf{A}_1^{-1}\mathsf{D}\mathsf{A}_2^{-1}
\end{pmatrix}.
\end{split}
\end{equation}

By employing the method based on the point transformation Eq.~\eqref{eqn:pointTransformationZ}, we can restrict ourselves to the case where we want to invert the matrix
\begin{equation}
\label{eqn:inverseGeneric}
\mathsf{P}=\begin{pmatrix}
\mathsf{S}&b\bm{a}\bm{f}_\alpha^T\\
b\bm{f}_\alpha\bm{a}^T&\mathds{I}+d\bm{f}_\alpha\bm{f}_\alpha^T
\end{pmatrix},
\end{equation}
where $\bm{f}_\alpha$ is a finite-norm vector with an infinite number of components, and $\bm{a}$ is a generic real vector whose dimension is the dimension of a row (column) of the generic square matrix $\mathsf{S}$.

By comparing Eq.~\eqref{eqn:inverseGeneric} and Eq.~\eqref{eqn:inversionFormula}, we compute:
\[
\left(\mathds{I}+d\bm{f}_\alpha\bm{f}_\alpha^T\right)^{-1}=\mathds{I}-\frac{d\bm{f}_\alpha\bm{f}_\alpha^T}{1+d\abs{\bm{f}_\alpha}^2},
\]
where we have used the Sherman-Morrison formula. Furthermore,
\[
\Tr[\bm{a}\bm{f}_\alpha^T\left(\mathds{I}-\frac{d\bm{f}_\alpha\bm{f}_\alpha^T}{1+d\abs{\bm{f}_\alpha}^2}\right)\bm{f}_\alpha\bm{a}^T\mathsf{S}^{-1}]=\frac{\abs{\bm{f}_\alpha}^2}{1+d\abs{\bm{f}_\alpha}^2}\bm{a}^T\mathsf{S}^{-1}\bm{a}.
\]

Therefore, by combining all the pieces together and using Eq.~\eqref{eqn:inversionFormula}, we obtain:
\begin{equation}
\label{eqn:inversionFormulaSpecific}
\mathsf{P}^{-1}=\begin{pmatrix}
\mathsf{S}^{-1}+b^2D^{-1}\abs{\bm{f}_\alpha}^2\mathsf{S}^{-1}\bm{a}\bm{a}^T\mathsf{S}^{-1}& -bD^{-1}\mathsf{S}^{-1}\bm{a}\bm{f}_\alpha^T\\
-bD^{-1}\bm{f}_\alpha\bm{a}^T\mathsf{S}^{-1}&\mathds{I}+[b^2\bm{a}^T\mathsf{S}^{-1}\bm{a}-d] D^{-1}\bm{f}_\alpha\bm{f}_\alpha^T
\end{pmatrix},
\end{equation}
with $D=1+\abs{\bm{f}_\alpha}^2(d-b^2\bm{a}^T\mathsf{S}^{-1}\bm{a})$. Eq.~\eqref{eqn:inversionFormulaSpecific} corresponds to Eq.~\eqref{eqn:inverseTwoQ}, and the specific cases analysed in Sec.~\ref{sec:coupling} are easily recognised.

If the impedance is multi-port, i.e. the external system is coupled thereto in different points (or equivalently, there is more than one resistor in the circuit), in general $\mathsf{D}$ in Eq.~\eqref{eqn:inversionFormula} is not a rank-one matrix anymore, and the formula introduced above does not hold. For these cases, which include all the possible situations we can face in a quantum circuit, an inversion formula still exists, although it is way more involved than Eq.~\eqref{eqn:inversionFormula}. It can be found in the Appendix D of Ref.~\cite{Parra-Rodriguez2018}.

\bibliography{Tutorial}

\end{document}